\pdfoutput=1
\documentclass[a4paper,fleqn,usenatbib,umoline]{mnras}

\usepackage{amsmath}  
\usepackage{amssymb}  
\usepackage{graphicx} 
\usepackage{txfonts}
\usepackage{ae,aecompl}
\usepackage{color}
\usepackage[multidot]{grffile}
\usepackage[capitalise]{cleveref}
\crefname{eq:}{Eq.}{Eqs.}
\crefname{fig:}{Fig}{Figs}
\usepackage{subfig}
\usepackage{umoline}
\usepackage{siunitx}
\usepackage{upgreek}
\usepackage[normalem]{ulem}

\newcommand{\be}{\begin{equation}}
\newcommand{\ee}{\end{equation}}
\newcommand{\bvec}[1]{\boldsymbol{#1}}
\newcommand{\tpartial}[1]{\frac{\partial\, #1}{\partial t}}
\newcommand{\cs}{c_\mathrm{s}}
\newcommand{\ceff}{c_\mathrm{eff}}
\newcommand{\sigmatot}{\sigma_{\mathrm{tot}}}
\newcommand{\rhoc}{\rho_\mathrm{c}}
\newcommand{\etaA}{\eta_{\mathrm{A}}}
\newcommand{\kms}{\mbox{km s$^{-1}$}}
\newcommand{\dr}[1]{\dfrac{d #1}{dr}}

\defcitealias{Rad2017}{H17}

\title[Gravitational instability of molecular clouds]{Gravitational instability of filamentary molecular clouds, including ambipolar diffusion; Non-isothermal filament}

\author[Hosseinirad et al.]{
Mohammad Hosseinirad,$^{1}$\thanks{E-mail: m.rad@birjand.ac.ir (MH); abbassi@um.ac.ir (SA); mroshan@um.ac.ir (MR);  naficy@birjand.ac.ir (KN)}
Shahram Abbassi, $^{2,3}$
Mahmood Roshan$^{2}$
\and Kazem Naficy$^{1}$
\\
 $^{1}$Department of Physics, University of Birjand, PO Box 615/97175, Birjand, South Khorasan, Iran\\
 $^{2}$Department of Physics, School of Sciences, Ferdowsi University of Mashhad, Mashhad, PO Box 91775-1436, Iran\\
 $^{3}$School of Astronomy, Institute for Studies in Theoretical Physics and Mathematics, PO Box 19395-5531, Tehran, Iran\\
}
\date{Accepted 2017 December 28. Received 2017 December 24; in original form 2017 November 28}

\pubyear{2017}

\begin{document}
\label{firstpage}
\pagerange{\pageref{firstpage}--\pageref{lastpage}}
\maketitle
\begin{abstract}
Recent observations of the filamentary molecular clouds show that their properties deviate from the isothermal equation of state. Theoretical investigations proposed that the logatropic and the polytropic equations of state with negative indexes can provide a better description for these filamentary structures. Here, we aim to compare the effects of these softer non-isothermal equation of states with their isothermal counterpart on the global gravitational instability of a filamentary molecular cloud. By incorporating the ambipolar diffusion, we use the non-ideal magnetohydrodynamics framework for a filament that is threaded by a uniform axial magnetic field. We perturb the fluid and obtain the dispersion relation both for the logatropic and polytropic equations of state by taking the effects of magnetic field and ambipolar diffusion into account. Our results suggest that, in absence of the magnetic field, a softer equation of state makes the system more prone to gravitational instability. We also observed that a moderate magnetic field is able to enhance the stability of the filament in a way that is sensitive to the equation of state in general. However, when the magnetic field is strong, this effect is suppressed and all the equations of state have almost the same stability properties. Moreover, we find that for all the considered equations of state, the ambipolar diffusion has destabilizing effects on the filament.
\end{abstract}


\section{Introduction}
It has been well established that the filamentary molecular clouds are the preferred birthplaces of stars \citep{2017arXiv171001030A}. Recent observations of the nearest Galactic molecular clouds (MCs) in the submillimeter wavelengths with \emph{Herschel Space Observatory} \citep{Pilbratt2010} has opened a new window to the understanding of the complex star formation process \citep{Andre-2010,Molinari-2010-ID667}. \emph{Herschel} shows the filaments are omnipresent in the cold interstellar medium (ISM). They are found both in star-forming \citep{Konyves2010,Bontemps2010} and non-star forming regions \citep{Menshchikov2010,Miville-Deschnes2010,Ward-Thompson2010}. This fact strengthens the idea that the filaments are the hosts of the early stages of formation of stars. Filaments are also pervasive in numerical simulations of MCs with various formation scenarios such as models in which gravity is the dominant agent and filaments formed by the global cloud collapse \citep[e.g.][]{Nagai1998,Burkert2004,Hartmann2007,Vazquez-Semadeni2007,Gomez2014,Wareing2016,Camacho2016} or models in which filaments are formed from the large-scale supersonic shocks \citep[e.g.][]{Klessen1998,Dib2007,Padoan2014,McKee-2007,Pudritz2013} and models in which the filaments are formed preferentially perpendicular to the magnetic field direction in a strongly magnetized turbulent cloud \citep[e.g.][]{Nakamura2008,Chen2014,Inutsuka2015,Federrath2016,Klassen2017}.

By taking a census of the filamentary structures in the IC5146 MC plus Aquila and Polaris regions in the Gould Belt, \citet{Arzoumanian2011} found strikingly that the filaments exhibit a narrow width distribution with a median value of $\sim 0.1$ pc. It should be noticed that \citet{Juvela2012} reported a larger width of $0.2 - 0.3$ pc for the filaments identified in the cold ISM regions previously found with the \emph{Planck} satellite. Likewise, \citet{Hennemann2012} found a range of $0.26 - 0.34$ pc for the massive gravitationally unstable filaments in DR21 ridge and filaments in Cygnus X \citep[see also][for a recent debate on the consistency of the existence of a characteristic filament width with the observed scale-free spatial power spectrum of the Herschel Polaris Flare image (at \SI{250}{\micro\meter}) \citealp{Miville-Deschenes2016}]{Panopoulou2017,2017arXiv171001030A}.

Another interesting feature of the identified filaments is that their radial profiles are somehow universal and can be described by a Plummer-like function of the form \citep{Plummer1911}
\begin{equation}
 \rho_p(r)=\dfrac{\rhoc}{[1+(r/R_{\text{flat}})^{2}]^{p/2}},
 \label{eq:PlummerLike}
\end{equation}
where $\rhoc$ is the central density, $R_{\text{flat}}$ is the radius of the flat inner region and $1.5<p<2.5$ is the exponent at large radii \citep{Arzoumanian2011,Juvela2012,Palmeirim2013}. They also showed that the dust temperature increases outward from the centre of the filaments. Taking these two parameters as $p=4$ and $R_{\text{flat}}=\sqrt{2\cs^2/(\pi G \rhoc)}$, \cref{eq:PlummerLike} will follow the density profile of an isothermal gas filament in the hydrostatic equilibrium \citep{Stod63,Ostriker64}. These facts might lead to the point that the filaments are not well described by the isothermal equation of state (IEOS), but instead might be better described by a cylinder with a polytropic equation of state with the polytropic exponent $\gamma_p<1$ \citep{Palmeirim2013,Toci2015}. In another work, \citet{Fischera2012} used a pressure-confined isothermal cylinder in equilibrium with the ambient medium to model the filamentary clouds in the IC5146 region. Another suggested explanation relies on a role that helical magnetic field can play in decrease of the steep slope of the isothermal profile \citep{FP2000I}. Recently, \citet{Recchi2013} proposed a thermal nature for the deviation of the density profile of the filaments from the IEOS. However, this will lead to a temperature about \SIrange{70}{170}{\kelvin} at a radius about 1 pc which is unlikely \citep{Toci2015}.

It is also reported by \citet{Arzoumanian2013} that for a sample of filaments in the Gould Belt, the molecular line observations show that these filaments can be divided into two subsets in terms of variation of their internal velocity dispersions with the column density. The first subset are gravitationally unbound and are thermally subcritical with a transonic total velocity dispersion ($\cs\lesssim \sigmatot< 2\cs)$ (where $\cs$ is the isothermal sound speed for a gas at $T=10$ K, corresponding to $\cs \simeq 0.2$ \kms) that show no meaningful relation with their measured column density while the second subset are gravitationally bound and are thermally supercritical with a total velocity dispersion that roughly depends on the column density as $\sigmatot \propto \Sigma^{0.5}$. Using a broad range of environments in the Galactic Plane that likely includes the filaments observed by \emph{Herschel}, however, \citet{Heyer2009} found that the velocity dispersion systematically varies with the surface density. In addition to this recent observation of the filaments, the outward increase of the velocity dispersion has been proven within GMCs \citep{Larson1981,Miesch1994} and also individual dense cores \citep{Fuller1992,Caselli1995}.

Many papers have been devoted to the theoretical study of the stability and the fragmentation of the filamentary MCs. In the pioneering work by \cite{Chandra53}, the stability of a homogeneous incompressible cylindrically symmetric gas was studied. They showed that a poloidal magnetic field can stabilize the filament. Other authors attempted to investigate this basic problem in  more practical sense \citep[e.g.][]{Stod63,Ostriker64b,Larson1985,Nagasawa87,Inutsuka1992,Fischera2012,Nakamura1993,Matsumoto94,Gehman1,Gehman2,Freundlich2014,Hanawa2015,Sadhukhan2016,Hanawa2017}.

Recently \citet[][hereafter H17]{Rad2017} studied the global gravitational instability of a magnetized filamentary cloud by carrying out linear perturbation analysis. They took into account the filament as a very long cylinder of the isothermal gas, threaded by a uniform poloidal magnetic field. Furthermore, they used the unperturbed magnetohydrodynamic (MHD) equations in the non-ideal framework, by incorporating the effect of ambipolar diffusion (AD) in the \emph{strong coupling approximation} (see \cref{sec:MHD}). They found that addition of the AD can destabilize the magnetized filament by increasing the growth rate of the most unstable mode. Additionally, they found that the AD leads to an enhancement of the fragmentation scale of the filament. They also showed that the system will proceed in this manner before it reaches to the state wherein no magnetic filed has been added.

The purpose of this paper is to complement \citetalias{Rad2017} by extending it to the non-isothermal equation of state (EOS). Our first candidate is the logatropic\footnote{Some authors called it logotropic.} equation of state (LEOS). This EOS is proposed by \citet{LizanoShu1989} for the first time to compromise between the theory and the observations that indicate the line width-determined velocity dispersion increases with the radius \citep[e.g.][but see also \citealp{Heyer2009}]{Larson1981}. Later, \citet{Gehman1,Gehman2} examined this EOS in a filamentary cloud as a proxy for the turbulence and performed a linear perturbation analysis. They demonstrated that using the LEOS can destabilize the filament considerably in comparison with the IEOS. Later on, by using a modified version, \citet{MP96} successfully fitted the velocity dispersions of both low- and high-mass cores derived from various MCs. After that, \citet{FP2000I} incorporated this modified version into the magnetohydrostatic equilibrium of a filamentary cloud which is pervaded by a helical magnetic field. They found that the magnetized filaments with the LEOS show shallower density profiles that fall off as $r^{-1}$ to $r^{-1.8}$ than those of magnetized filaments with the IEOS for which the density profiles fall off as $r^{-1.8}$ to $r^{-2}$. The second candidate is the polytropic equation of state (PEOS). As mentioned earlier, observations suggest that the PEOS seems to be a better choice for the modeling of the filaments. Recently, \citet{Freundlich2014} made use of the local stability analysis and showed that a filament with the PEOS is more stable than its isothermal counterpart. Following it, \citet{Sadhukhan2016} added the magnetic field to this problem. Moreover, \citet{Toci2015} demonstrated that the filaments with non-isentropic pressure support, are stable against radial collapse in the observed range of axis-to-surface density contrast.We are encouraged by this findings to extend our analysis to the PEOS. We aim this paper can shed some insight into this problem.

The paper is structured as follows. In \cref{sec:MHD} we describe the non-ideal MHD equations. The EOSs, physical parameters and unperturbed state are introduced in \cref{sec:EOS,sec:parameters,sec:unperturbed} respectively. We linearize the non-ideal MHD equations in \cref{sec:perturbed}. The boundary conditions and the numerical method are outlined in \cref{sec:boundary,sec:nemeric} respectively. The results are given in \cref{sec:results}. Finally, we conclude and summarize our results in \cref{sec:conclusion}.
\section{Basics}
\subsection{MHD equations with AD}\label{sec:MHD}
In partially ionized media such as interstellar medium, MCs and protoplanetary discs, MHD equations must be modified to account for non-ideal MHD effects namely ohmic dissipation, Hall effect and ambipolar diffusion, according to the degree of ionization and the strength of magnetic field. In the literature, two approaches are usually exploited for formulation of non-ideal MHD \citep[see][for a review]{Zweibel2015}. In the first approach the tensor conductivity is used to replace the current density in the induction equation. This method is specially useful when there are several types of charge carriers \citep[see e.g.][]{Cowling56,Norman85,NakanoUmebayashi86,WardleNg99,Salmeron2003,Wardle2007,Zhao2016,Wurster2016,Masson2016}. In the second one, the fluid equations for different charge species are used as a start point. This multi-fluid formulation, can be simplified to a single coupled ion-neutral fluid form, when the degree of ionization is low enough. This will happen if the gravitational force and pressure gradient of the ions can be neglected in comparison with the frictional and Lorentz forces\footnote{Electrons contribution to the momentum exchange is negligible and ignored.}\citep[see e.g.][]{MacLow95,BalbusTerquem2001,Oishi2006,Choi2009,Gressel2015,Ntormousi-2016}. In what follows we make use of this so-called \emph{strong coupling approximation} \citep{Shu83}. Our set of MHD equations with the self-gravity and the AD term are the continuity equation,
\begin{equation}
  \tpartial{\rho} + \nabla\cdot\left(\rho\bvec{u}\right) = 0,
  \label{eq:cont}
\end{equation}
the equation of motion,
\begin{equation}
  \rho\tpartial{\bvec{u}} + \rho\left(\bvec{u}\cdot\nabla\right) \bvec{u}
  + \nabla p + \rho\nabla\psi
  - \dfrac{1}{4\pi}\left(\nabla\times\bvec{B}\right)\times\bvec{B} = 0,
  \label{eq:force}
\end{equation}
the induction equation,
\begin{equation}
\tpartial{\bvec{B}} + \nabla\times\left(\bvec{B}\times\bvec{u}\right) - 
\nabla\times\Bigg\{\bigg[\etaA\left(\nabla\times\bvec{B}\right)\times\bvec{B}\bigg]\times\bvec{B}\Bigg\}= 0,
\label{eq:ind}
\end{equation}
and the Poisson equation for gravity,
\begin{equation}
  \nabla^{2}\psi = 4\pi G \rho.
  \label{eq:poisson}
\end{equation}
In these equations, $\rho$ is the neutral gas density, $\bvec{u}$ is the fluid velocity, $p$ is the pressure, $\psi$ is the gravitational potential and $\bvec{B}$ is the magnetic field strength, where $\etaA$ is the AD coefficient.
\begin{figure*}
  \centering
    \includegraphics[scale=0.285]{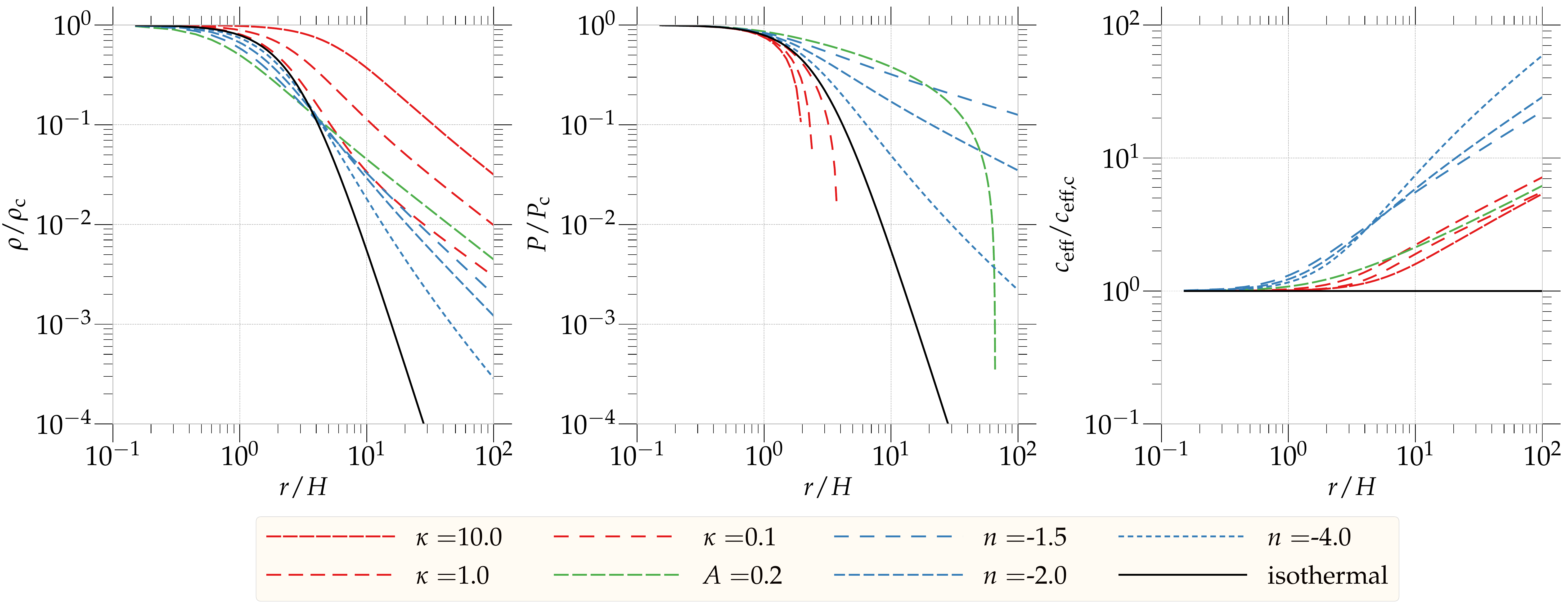}
    \caption{From left to right, the density, the pressure and the effective sound speed profiles of the IEOS, MPEOS, GEOS ($\kappa =$ 0.1, 1, 10) and PEOS ($n$ = -1.5, -2 and -4).}
    \label{fig:RhoPressureSigmaLogLog}
\end{figure*}
\subsection{Equation of state}\label{sec:EOS}
To complete our set of \cref{eq:cont,eq:force,eq:ind,eq:poisson} we need a prescription for the pressure. In the \citetalias{Rad2017}, we used the simplest case i.e. the IEOS for describing the equilibrium state of a filament of gas in cylindrical coordinates. Here we extend our analysis in \citetalias{Rad2017} to include more physically plausible equations of state. More specifically we use two other types of barotropic EOS, i.e. the LEOS and the PEOS. We formulate the LOES in two different ways which have already been used in the relevant literature
\begin{equation}
 p = c_s^2\,\rho + p_0 \log\,(\rho/\rhoc)
 \label{eq:GEOS}
\end{equation}
\citep[][but see also \citealp{LizanoShu1989}]{Gehman1,Gehman2} and
\begin{equation}
 p = p_c[1 + A \log\,(\rho/\rhoc)]
 \label{eq:MPEOS}
\end{equation}
\citep{MP96,FP2000I}
where $\rhoc$ is the density at the filament axis, $\cs$ is the isothermal sound speed and $p_0$ and $A$ are empirical constants. \cite{Gehman2} suggested that $10<p_0<70$ picodynes cm$^{-2}$ is an acceptable range for MCs. By analyzing molecular cloud cores, \citet{MP96} found $A\simeq 0.2$. Moreover, \cite{FP2000I} used this value in \cref{eq:MPEOS} to construct the magnetohydrostatic equilibrium for the filamentary clouds. We refer to \cref{eq:GEOS} and \cref{eq:MPEOS} as the GEOS and the MPEOS respectively.

On the other hand, in the case of PEOS, we use the following form
\begin{equation}
 p = p_c\,(\rho/\rhoc)^{\gamma_p},
 \label{eq:PEOS}
\end{equation}
where $\gamma_p$ is the polytropic exponent. We assume $\gamma_p$ to be the same as the adiabatic exponent $\gamma_A$, i.e. the filament is isentropic. This means during a density perturbation, entropy remains both spatially and temporally constant. Observations of GMCs, filamentary clouds and individual dense cores, put forward a family of PEOS for which $0<\gamma_p<1$. Since, it is common to define $\gamma_p = 1 + 1/n$, this will correspond to $-\infty<n<-1$\citep{Viala1974,Maloney1988}.
\subsection{Physical parameters}\label{sec:parameters}
It is obvious that in the \emph{strong coupling approximation}, the effect of AD is appeared only in the induction equation where the new introduced term determines the amount of AD. Considering $\gamma$ as the drag coefficient in ion-neutral collisions and $\rho_i$ as the ion density, the AD coefficient can be expressed as
\begin{equation}
 \etaA = \dfrac{1}{4\pi \gamma \rho_i \rho},
 \label{eq:eta}
\end{equation}
\citep{Shu1987}. The ion-neutral drag coefficient is 
\begin{equation}\label{eq:gamma}
 \gamma = \dfrac{\langle \sigma w \rangle}{m_i + \mu},
\end{equation}
where $m_i$ and $\mu$ are the ion mass and mean molecular weight both per hydrogen atom mass that are taken to be about $30\ m_H$ and 2.36 respectively. $\langle \sigma w \rangle$ is the ion-neutral coupling coefficient that is taken to be $1.9 \times 10^{-9}$ cm$^3$ s$^{-1}$ \citep{Draine83}. Having considered foregoing quantities the estimated values for $\gamma$ is $3.5 \times 10^{13}$ cm$^3$ g$^{-1}$ s$^{-1}$. To fully determine $\etaA$, we also need to know the ion density. Following \citet{Elmegreen79}, it is assumed  that due to the cosmic radiation, one can approximate ion density in MCs as
\begin{equation}\label{eq:ion-density}
 \rho_i = C \rho^{1/2},
\end{equation} where the constant $C$ is $3 \times 10^{-16}$ cm$^{-3/2}$ g$^{1/2}$. Substituting \cref{eq:ion-density} in \cref{eq:eta} and considering $\gamma C$ as a new parameter $\alpha$ will give us 
\begin{equation}
 \etaA = \dfrac{1}{4\pi \alpha \rho^{3/2}}.
 \label{eq:eta_A_with_alpha}
\end{equation}
With these characteristics, a typical MC with density of $4 \times 10^{-20}$g cm$^{-3/2}$ will have the fractional ionization of $\sim 10^{-7}$ which seems consonant with values of $10^{-6}$ to $10^{-8}$ that are come from observation.
\subsection{Unperturbed state}\label{sec:unperturbed}
The unperturbed filament is considered to be in hydrostatic equilibrium. We use standard cylindrical coordinates ($r, \phi, z$) with the origin at the filament centre. The filament is supposed to be very long. An initial uniform magnetic field $\bvec{B_0} = B_0\,\hat{z}$ threads the filament, so it does not contribute in supporting the filament against its self-gravity. Having such a configuration, the momentum equation \cref{eq:force} and the Poisson equation \cref{eq:poisson} can be combined and solved to determine the density profile at the equilibrium state. For the isothermal filament a well-known analytical solution exists as
\begin{equation}
 \rho(r)=\rhoc (1+\dfrac{r^2}{8H^2})^{-2}
 \label{eq:rho}
\end{equation}
where $\rhoc$ is the central density \citep{Stod63,Ostriker64}. $H$ is a radial scale length which is defined as
\begin{equation}
 H = \dfrac{c_s}{\sqrt{4\pi G \rhoc}}
\end{equation}
where $\cs$ is the isothermal sound speed and $G$ is the gravitational constant. For a filament with the temperature of 10 K or equivalently the thermal sound speed of 0.2 \kms and a central density of $4\times 10^{-20}$g cm$^{-3}$, $H$ will be $\approx 0.035$ pc. For the GEOS, MPEOS and the negative index PEOS, the analytical solution does not exist. Determining the initial values is indeed the main obstacle in the way of computing the density of  \cref{eq:force,eq:poisson} \citep[e.g.][]{Gehman1} which can be solved numerically. This were done using \texttt{odeint} routine from the \texttt{Scipy} library \citep{scipy}.

\cref{fig:RhoPressureSigmaLogLog} demonstrates the density, the pressure and the effective sound speed of the above-stated EOSs. We use dimensionless quantities as described in \cref{sec:AD_coef}. In the GEOS, we set the dimensionless free parameter $\kappa$ = 0.1, 1 or 10. \citet{Gehman2} pointed out $6<\kappa<50$ matches the observation. For the MPEOS we set the dimensionless parameter $A=0.2$, as suggested by \citet{MP96}. For the PEOS, we choose the polytropic indexes as $n$ = -1.5, -2 and -4 ($\gamma_p = 1+1/n$ = 1/3, 1/2 and 3/4). This encompasses the observed range of filaments in the IC5146 \citep{Arzoumanian2011,Toci2015}. The density and pressure are normalized by their central values. The left-hand panel depicts the density profiles. Comparing with the isothermal filament, ones with the GEOS have larger density all over the radial extent. Filaments with the MPEOS and PEOSs are slightly more concentrated at the centre but fall off more slowly at larger radii. The middle panel shows the pressure profile for each EOS. It should be noted that for the GEOS, there is a cut-off radius at which the pressure becomes negative. This cut-off radius is very near the centre of the filament and takes smaller values for larger $\kappa$. This is also the case for the MPEOS but at very larger radius. However, filaments with the PEOS do not exhibit this characteristic. Their pressure asymptotically approach to the zero, but they have larger pressure all over the way with respect to the isothermal filament. In the right-hand panel, we illustrate the effective sound speed $\ceff=(dp/d\rho)^{1/2}$ which is crucial for estimating the length-scale of fragmentation i.e. the Jeans length, $\lambda_J=\ceff(\pi/G\rho)^{1/2}$. This panel shows that the effective sound speed increases monotonically with radius in all filaments with the GEOS, MPEOS and PEOS.
\subsection{The Linearized non-ideal MHD equations}\label{sec:perturbed}
In this section, we perform global perturbation analysis of the governing non-ideal MHD equations in the presence of self-gravity, \cref{eq:cont,eq:force,eq:ind,eq:poisson}. Perturbing these equations in dimensionless form (see \cref{sec:AD_coef}) to the first order gives
\be
    \tpartial{\rho_{1}} + \nabla\rho_0\cdotp\bvec{u}_1 + \rho_0\nabla\cdotp\bvec{u}_1=0,\label{eq:cont1}
  \ee
  \be
    \rho_{0}\tpartial{\bvec{u}_{1}} + \nabla p_{1} + \rho_{0}\nabla\psi_{1}
    + \rho_{1}\nabla\psi_{0} - \left(\nabla\times\bvec{B}_1\right)\times\bvec{B}_0 = 0,\label{eq:mom}
  \ee
  \begin{equation}
\tpartial{\bvec{B}_{1}}
+ \nabla\times\left(\bvec{B}_{0}\times\bvec{u}_{1}\right) 
-\etaA\nabla\times\Bigg\{\bigg [\left(\nabla\times\bvec{B}_1\right)\times\bvec{B}_0\bigg ]\times\bvec{B}_0\Bigg\} = 0.
\label{eq:ind1}
\end{equation}
  \be
    \nabla^{2}\psi_{1} = \rho_{1}.
  \ee 
Here, the subscripts ``0'' and ``1'' are reserved for the unperturbed and perturbed quantities. It should also be emphasized that meanwhile of linearization $\etaA$ is taken to be constant. This simplifies the calculations as well as interpretation of the results. The barotropic form of EOSs, let us to linearize them as
\be p_1=\dfrac{d P}{d\rho}(\rho_0)\rho_1 \equiv P^{\prime}(\rho_0)\rho_1. \ee
In the following, for the sake of simplicity, we restrict ourselves to the propagation of axisymmetric perturbations. In this case all the perturbations can be expressed as superposition of their axisymmetric Fourier modes. Furthermore we investigate the perturbations which propagate only along the axis of the filament, i.e. the $z$ axis. Therefore, the general form for the Fourier mode for this type of perturbations reads as \begin{equation}
  \left(
  \begin{array}{c}
    \rho_{1}(\bvec{x}, t) \\
    \bvec{u}_{1}(\bvec{x}, t) \\
    \bvec{B}_{1}(\bvec{x}, t) \\
    \psi_{1}(\bvec{x}, t)
  \end{array}\right) =
  \Re \left[ ~ \left(
  \begin{array}{c}
    f(r) \\
    \bvec{v}(r) \\
    \bvec{b}(r) \\
    \phi(r)
  \end{array}\right) \exp{(ikz-i\omega t)} \right],
  \label{eq:rho1u1B1psi1}
\end{equation}
where $\omega$ is the angular frequency, $k$ is the vertical wave number and $\Re$ denotes to the real part. Now we substitute these Fourier modes to the linearised equations \eqref{eq:cont1}-\eqref{eq:ind1}. We simplify this set of equations while restricting ourselves to the unstable modes which grow with time for which $i\omega$ is a real negative number. If we substitute $-\Omega$ for $i\omega$ and $w$ for $i\omega v_r$ the resultant equations read
\begin{align}
-\rho_0 w + & fP''\dr{\rho_0} + P'\dr{f}+\rho_0\dr{\phi}+f\dr{\psi_0} \nonumber \\
                - & (\frac{B_0^3k^2\etaA}{\etaA B_0^2k^2+\Omega})\dr{b_z}-\frac{B_0^2k^2w}{\Omega(\etaA B_0^2k^2+\Omega)} \nonumber \\ 
                + & B_0\dr{b_z}=0,\label{eq:finalODE2}
\end{align}
\be r\rho_0\dr{w}+\rho_0w+r(-\Omega^2-k^2P')f-rk^2\rho_0\phi+rw\dr{\rho_0}=0,\label{eq:finalODE3}\ee
\begin{align} 
-\etaA & B_0^2\bigg(b_zk^2r-r(\frac{d^2b_z}{dr^2})-\dr{b_z}\bigg)-\Omega b_z r \nonumber \\
      - & B_0r\bigg(\frac{-\Omega f}{\rho_0}-k^2\frac{\phi}{\Omega}-k^2\frac{fP'}{\Omega\rho_0}+\frac{w}{\Omega\rho_0}\dr{\rho_0}\bigg)=0 \label{eq:finalODE1}\end{align}
and
\be r\frac{d^2\phi}{dr^2}+\frac{d\phi}{dr}-rk^2\phi-rf=0.\label{eq:finalODE4}\ee
\begin{figure*}
  \includegraphics[scale=0.5]{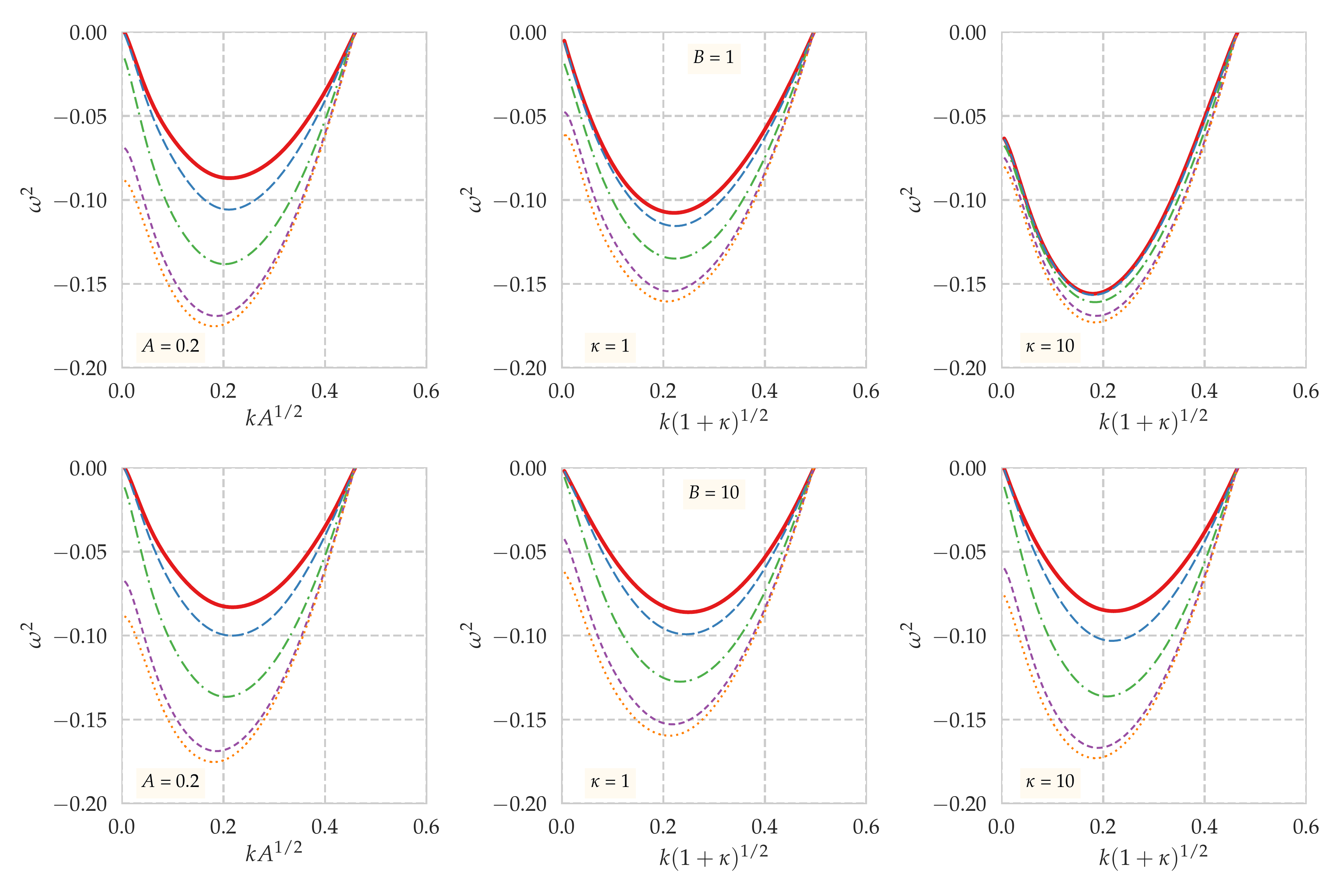}
    \caption{Dispersion relation of the filament with the LEOS. The left column belongs to the MPEOS, and the middle and right columns belong to the GEOS. Upper and lower panels show the dispersion relation when $B = 1$ and $B = 10$ respectively. In the left-hand panel $A =$ 0.2 and in the middle and right-hand panels $\kappa =$ 1 and $\kappa =$ 10 respectively. In each panel the horizontal axis is the wave number $k$ and the vertical axis is $\omega^2$ that are normalized in the units of $(4\pi G \rhoc)^{1/2}/\cs$ and $4\pi G \rhoc$ respectively. The wave numbers are multiplied by $A^{1/2}$ or $(1+\kappa)^{1/2}$ to account for the usage of thermal sound speed as the velocity unit. The solid red line represents dispersion relation for the case in which $\etaA=0$. Other lines demonstrate different $\etaA$ values from top to bottom as $1$ (blue long dashed), $10$ (green dash-dotted), $10^2$ (violet short dashed) and $10^3$ (orange dotted).}
    \label{fig:mplogaOmega2k}
\end{figure*}
\subsection{Boundary conditions}\label{sec:boundary}
\cref{eq:finalODE2,eq:finalODE3,eq:finalODE1,eq:finalODE4} constitute a system of coupled ordinary differential equations (ODEs) that must meet, in total, seven boundary conditions (BCs) at the centre of the filament and infinity. Due to the axial symmetry of the perturbations, all the radial force components, as well as the radial velocity, must vanish at the filament centre. Moreover, all the perturbations and their derivatives must vanishe at the infinity. The linear ODE system, leaves also the freedom of choosing all dependent variable but one and then solve for the other variables. Considering all the above conditions, we choose BCs as \begin{align}
 f&=1,\quad \frac{d\phi}{dr}=0,\quad w=0,\quad \frac{db_z}{dr}=0 \quad at\quad r=0.
 \label{boundary-0}
\end{align}
\begin{align}
 f&=0,\quad w=0, \quad \frac{db_z}{dr}=0 \quad at \quad r=\infty.
 \label{boundary-inf}
\end{align}
\subsection{Numerical methods}\label{sec:nemeric}
Having determined BCs, \cref{eq:finalODE2,eq:finalODE3,eq:finalODE1,eq:finalODE4} can be solved simultaneously. To do so, we take into account $k$ as the eigenvalue \footnote{The ODE system under consideration is actually a disguised eigensystem.} and $\Omega$ as a parameter which is initialized before calculation. We use a Newton-Raphson-Kantorovich (\texttt{NRK}) relaxation algorithm \citep{Garaud-thesis} that takes the advantage of the second order finite-difference discretization over a mesh. This algorithm indeed convert the ODE system to an algebraic system of equations. We use 2000 equally spaced mesh points throughout the calculation. We choose $r = 50$ as the effective infinity, however, the values of the eigenfunctions at the large radii, sometimes enforce a larger or smaller value for the effective infinity chosen as $r = 300$ and $r = 25$ respectively. The \texttt{NRK} algorithm needs an initial guess to start. At the first, when the AD and the magnetic field are not present, using a reasonable initial guess will readily make the system to converge. We use this result as an initial guess when magnetic field is present. The appropriate initial guess when the AD is present is taken from the nearest previous solution.
\begin{figure*}
  \includegraphics[scale=0.5]{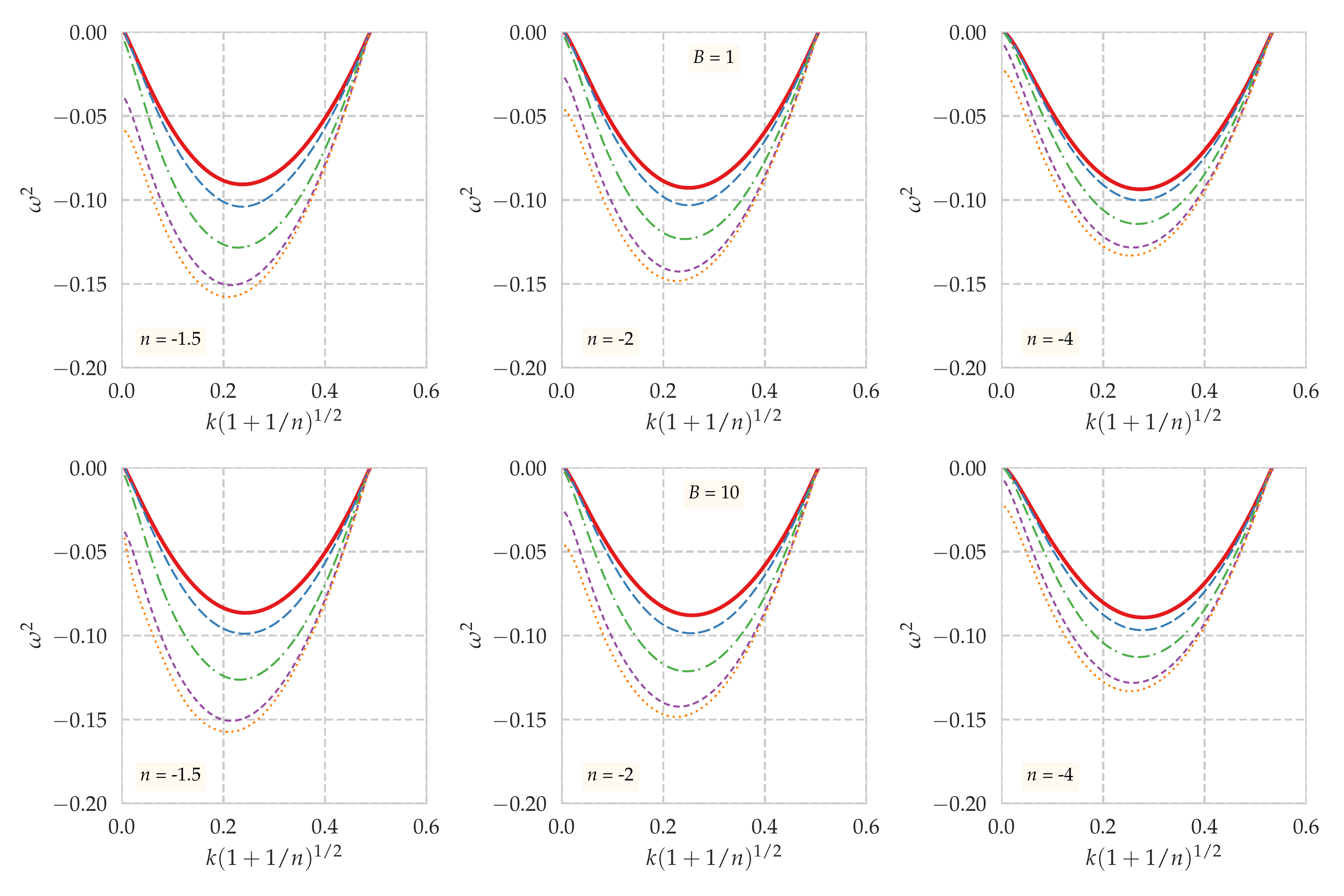}
    \caption{Same as the \cref{fig:mplogaOmega2k} but for the polytropic filament. Panels from left to right have polytropic indexes of $n$ = -1.5, -2 and -4 respectively. The wave numbers are multiplied by $(1+1/n)^{1/2}$ to account for the usage of thermal sound speed as the velocity unit.}
    \label{fig:polyOmega2k}
\end{figure*}
\section{Results}\label{sec:results}
Based on the aforementioned numerical method, we search for the $\omega$ values for which a solution exists in order to determine the dispersion relation. \citetalias{Rad2017} determined the dispersion relation of an isothermal filament threaded by a uniform axial magnetic field in the presence of AD. It showed that for the IEOS, the effect of magnetic field in the weak regime ($B = 0.1$) on the shape of dispersion relation can be ignored, even in a very strong AD regime ($\etaA = 10^4$). 

\cref{fig:mplogaOmega2k} shows the dispersion relations for two flavors of the LEOS, namely the MPEOS (left-hand panel) and the GEOS (middle and right-hand) (see \cref{sec:EOS}). For the MPEOS, $A = 0.2$, while for the GEOS, $\kappa = 1$ or $\kappa = 10$. In the top and bottom panels, the magnetic field strength is $B = 1$ and $B = 10$ respectively. Each panel demonstrates dispersion relations in different AD regimes, from $\etaA = 1$ to $\etaA = 10^3$. We found that the $\etaA < 1$ has not any significant effect on the dispersion relation even in a very strong magnetic field regime. We also found that for $\etaA > 10^3$, the dispersion relation coincides, effectively, with a system in which the magnetic field is zero. In other words, systems with large values of $\etaA$, respond against global perturbations in a way that is very similar to the systems which do not include magnetic field.

One should note that the wavenumber $k$ is scaled by the factors $A^{1/2}$ and $(1+\kappa)^{1/2}$ in the case of the MPEOS and the GEOS respectively which is indeed the effective sound speed $(dp/d\rho)^{1/2}$ at the centre. This scaling is done because in the scale length definition we have used the isothermal sound speed $\cs$.

In order to investigate the stability of the filament in the pure Jeans mode, i.e. without magnetic field ($B = 0$), one can analogously think about the dispersion relation in the strong AD regime, because the AD counteracts the effect of magnetic field \citepalias{Rad2017}. Looking at the dispersion relations in \cref{fig:mplogaOmega2k} when $\etaA = 10^3$ (the lowest curve in all panels), it is easy to see that the response of the filament to the perturbation for the MPEOS is very similar to the GEOS for $\kappa = 10$. In the case of $\kappa = 1$, the dispersion relation has also almost the same shape, but its critical wavelength and its fastest growing mode (i.e. one with the largest $|\omega^2|$ ) are both a little smaller than those of the GEOS with $\kappa = 10$ and the MPEOS.

Now let us first explain the top panels, where the magnetic field strength is $B = 1$, with more details. In this case, when the AD is gradually reduced, the magnetic field gradually becomes more effective to stabilize the filament. This is completely visible for the MPEOS where the magnetic field is able to decrease the growth rate of the fastest growing mode about 50 per cent. The efficiency of magnetic field in reinforcing the stability of the filament, is decreased for the GEOS with $\kappa = 1$. For $\kappa = 10$, the magnetic field becomes totally inefficient to stabilize the filament. Reported by \citetalias{Rad2017} and \citet{Gehman2}, there exist an upper limit for the magnetic field strength at which the stability of the filament is no longer increased. This saturation limit depends on the EOS. To check it, the computation of dispersion relation is also done for more powerful magnetic field strength of $B = 10$. Comparing the bottom panels of \cref{fig:mplogaOmega2k} with the top panels, one can find that the filament with the MPEOS as well as the GEOS with $\kappa = 1$ are already saturated by the magnetic field strength of $B = 1$. This is not the case for the GEOS with $\kappa = 10$. Also, as we already mentioned, the critical wavelength is independent from the magnetic field strength and is almost the same in these three LEOSs, but smaller when $\kappa = 1$.

It is important mentioning that, it is clear from \cref{fig:mplogaOmega2k,fig:polyOmega2k}, that changing the magnetic field as well as the AD coefficient does not influence the instability interval. In other words, the critical wavelength, i.e. the smallest unstable wavelength, does not depend on $B$ and $\etaA$. Analytically, one can show that when $\omega = 0$ , the magnetic field and the AD coefficient, are factored out from \cref{eq:finalODE2,eq:finalODE3,eq:finalODE1}. However, the AD can effectively change the growth rate of the perturbations. On the other hand the $\kappa$ parameter, substantially shortens the instability interval. It can be easily understood in the sense that $\kappa$ is a representative for the pressure. Therefore, by increasing $\kappa$, the pressure budget of the system increased, and naturally the stabilizing effects of the pressure suppress the small wavelengths to grow.

The dispersion relations of three polytropic indexes of $n = -1.5$, $n = -2$ and $n = -4$ ($\gamma_p = 1/3, 1/2$ and $3/4$), are displayed by \cref{fig:polyOmega2k}. The horizontal axes are again scaled by the effective sound speed which is $(1+1/n)^{1/2}$. Regarding the effect of magnetic field and AD, \cref{fig:polyOmega2k} at a glance suggests that the general behaviour of the dispersion relation for a filament with the PEOS, is similar to the MPEOS and the GEOS. There is not a meaningful difference between the top and bottom panels, suggesting that in terms of the stability, filaments with these three PEOSs, almost have been saturated by a magnetic field strength of $B = 1$. In the strongest AD regime ($\etaA=10^3$) that the magnetic field has the least effect on the stability of the filament, one with $n=-1.5$ that is shown in the left-hand panel, has the fastest growth rate and also the largest critical wavelength. Decreasing $n$, reduces both the fastest growth rate and the critical wavelength (middle and left-hand panels). Comparing this figure with the figure 2 in \citetalias{Rad2017}, one can realize that among the six EOSs that we computed their dispersion relations, the PEOS with $n=-4$ is the most similar filament to the one with the IEOS in terms of the gravitational instability.
\begin{figure}
  \centering
    \includegraphics[scale=0.45]{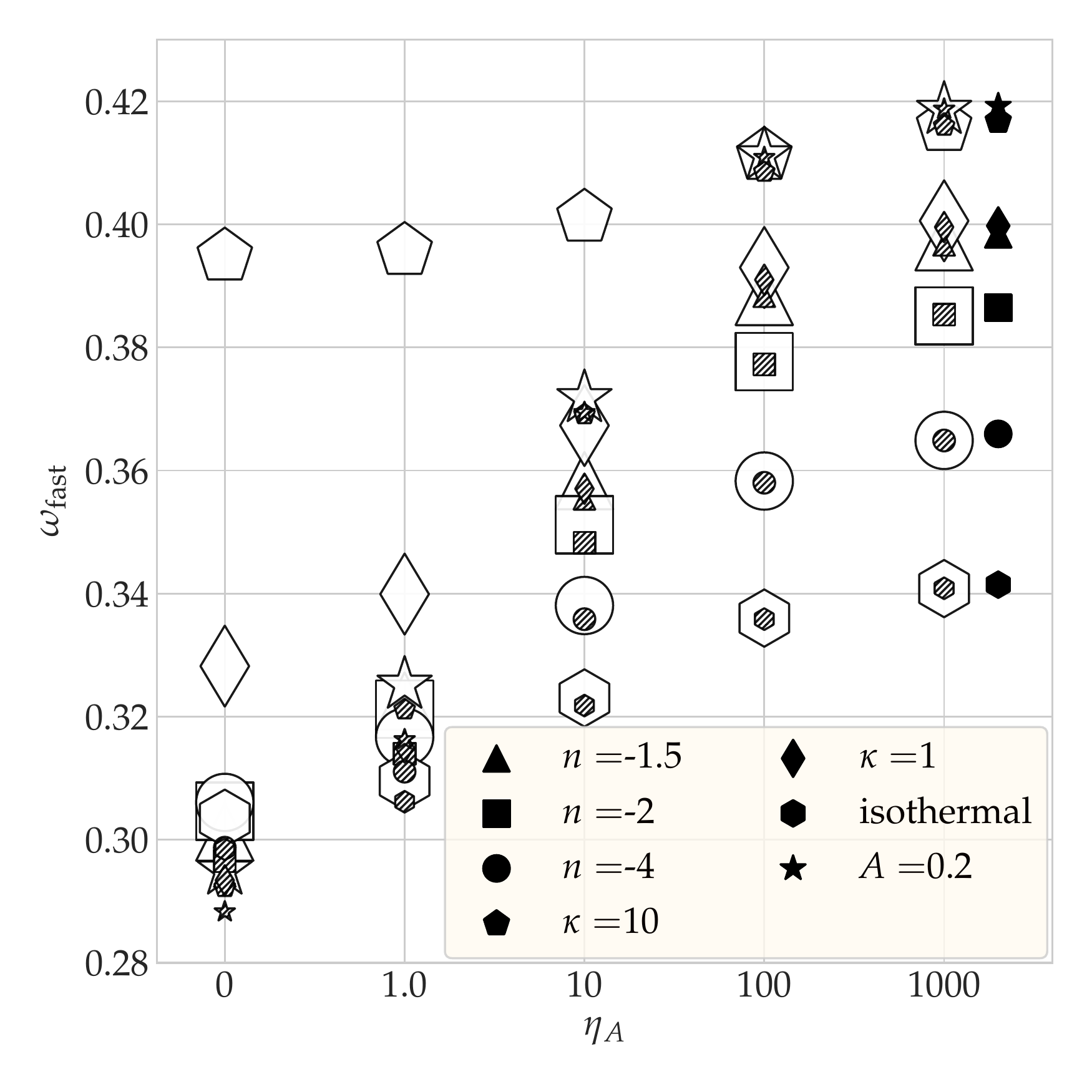}
    \caption{The fastest growth rate versus $\etaA$. Large open and small hatched markers show $\omega_{\text{fast}}$ for $B = 1$ and $B = 10$ respectively. The filled black markers show $\omega_{\text{fast}}$ when no magnetic field is present which are drawn next to the $\etaA = 1000$ for comparison.}
    \label{fig:OmeFastEta}
\end{figure}

The fastest growing mode, plays the key role in the fragmentation process. To better understand the fragmentation of the filament, we continue with the investigation of the dominant mode in more details. \cref{fig:OmeFastEta} shows the growth rate $\omega_{\text{fast}} = \sqrt{|\omega^2|}$ of the perturbations from weak to strong AD regimes for various EOSs and two magnetic field strength $B=1$ (large open markers) and $B=10$ (small hatched markers). Also, the case of pure Jeans instability is shown by filled black markers. From this figure, one can see that in the pure Jeans regime, filaments with the LEOS have larger $\omega_{\text{fast}}$ than others which is reasonable, because they are supported by lower gas pressure against their self-gravity (see \cref{fig:RhoPressureSigmaLogLog}). Among LEOSs, the MPEOS has the largest growth rate and the next ones are GEOSs with $\kappa=10$ and $\kappa=1$ respectively. They are followed by $n=-1.5$, $n=-2$ and $n=-4$ until the IEOS which has the smallest growth rate. In the presence of a magnetic field of $B=1$, the GEOS with $\kappa = 10$ has the largest growth rate. With a noticeable difference the next one is the GEOS with $\kappa=1$. The difference between the PEOSs is little and they all have smaller growth rates than the GEOSs. Here the MPEOS has the smallest growth rate. By increasing the AD coefficient $\etaA$, the above-mentioned gap between $\kappa=10$ and $\kappa=1$ becomes smaller. Also, it is obvious that with one notable exception the ordering in vertical direction is preserved. The exception is the MPEOS which by increasing $\etaA$, it's growth rate substantially increased insofar becomes the largest one. Moreover, similar to \citetalias{Rad2017}, one can immediately recognize that in the presence of the magnetic field, increasing $\etaA$, leads the stability properties of the system to be more similar to the pure Jeans case. When the filament is threaded by the stronger magnetic field $B=10$, the overall picture remains the same as $B=1$, specially in the strong AD and the pure Jeans regime, but the magnetic field is now more capable to suppress the instability for all the EOSs.

\begin{figure}
  \centering
    \includegraphics[scale=0.45]{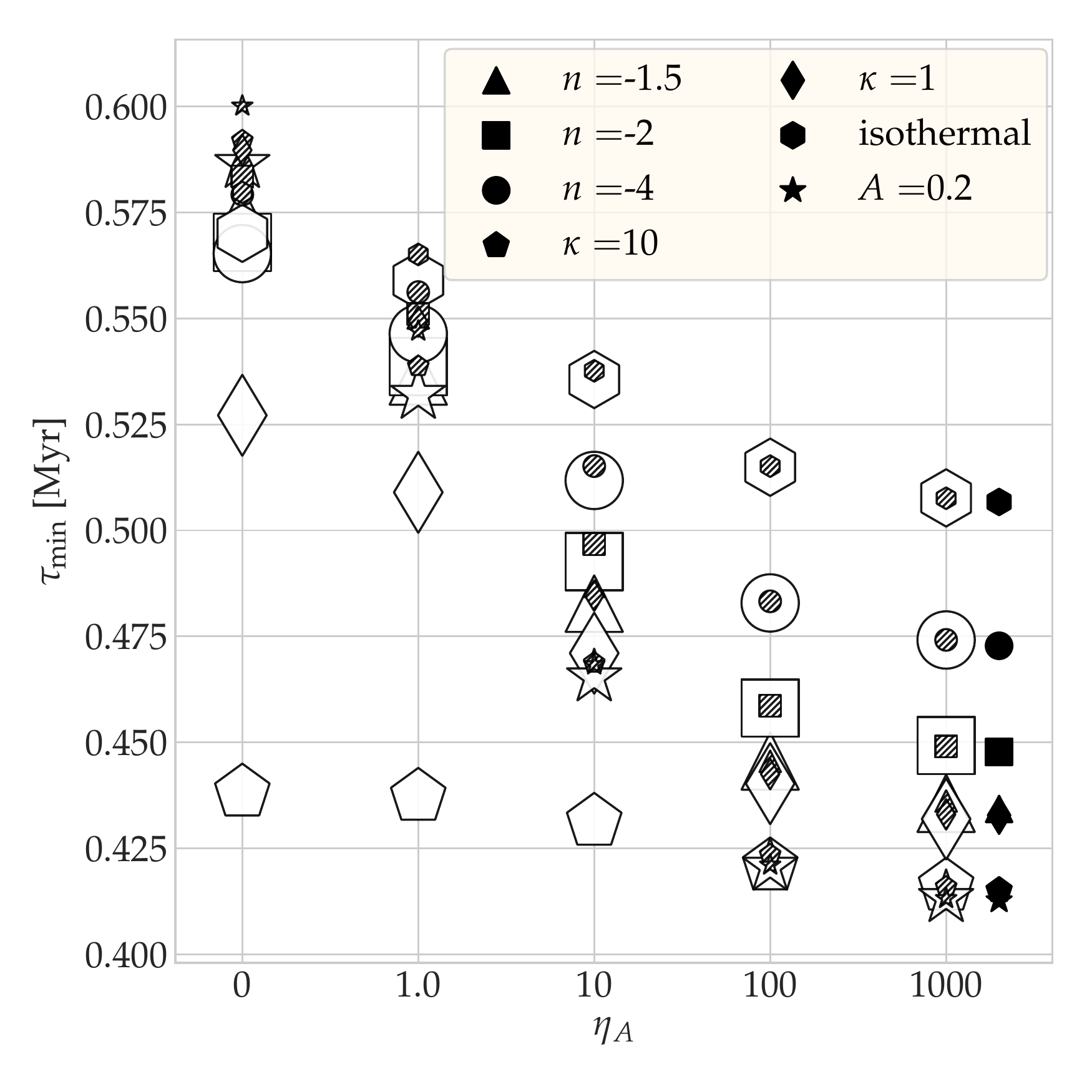}
    \caption{Same as the \cref{fig:OmeFastEta}, but for the minimum time needed for the fragmentation.}
    \label{fig:TimeMinEta}
\end{figure}
\cref{fig:TimeMinEta} illustrates the e-folding growth time of the perturbation. We take it into account as a representative for the minimum time needed for the fragmentation which is calculated as $\tau_{\text{min}}=1/\omega_{\text{fast}}$. All the above-mentioned details respecting $\omega_{\text{fast}}$ can be repeated, but certainly in an inverse picture. For the IEOS without effect of magnetic field, $\tau_{\text{min}}\approx 0.51$ Myr. All other EOSs have shorter fragmentation time-scales, with the minimum at $\approx 0.41$ Myr which belongs to the MPEOS. Turning on the magnetic fields $B=1$ and $B=10$, increases all the fragmentation time-scales at the most $\approx 0.59$ Myr and $\approx 0.6$ Myr for the MPEOS, respectively. It should be noted that, all fragmentation time-scale experience a reduction by adding AD.

\begin{figure}
  \centering
    \includegraphics[scale=0.45]{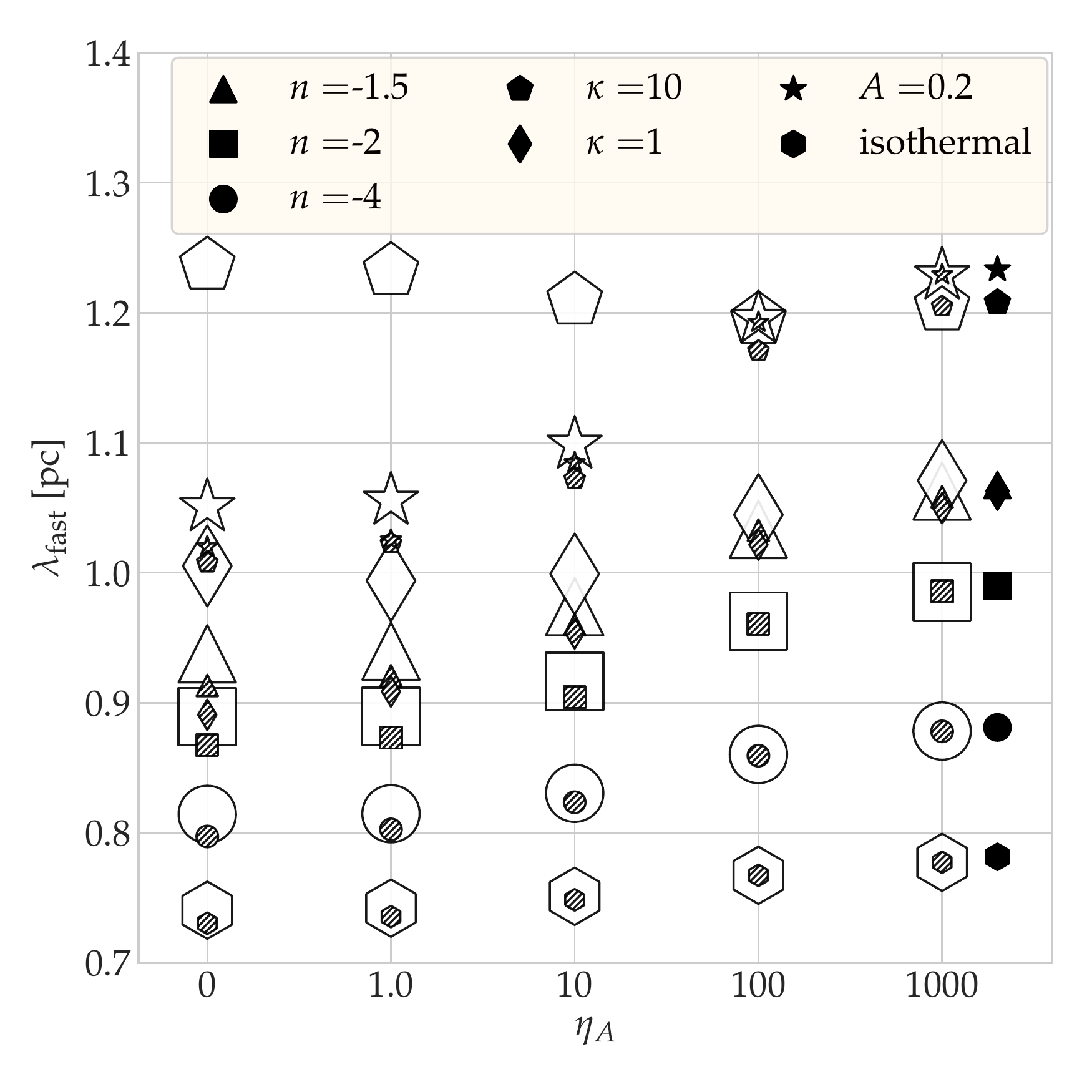}
    \caption{Same as the \cref{fig:OmeFastEta}, but for the length-scale of the fragmentation.}
    \label{fig:LambdaMinEta}
\end{figure}

By looking at \cref{fig:LambdaMinEta}  we can vividly realise that, the wavelength that correspond to the fastest growing mode is compared for a given EOS, magnetic field strength and AD coefficient. It is computed as $\lambda_{\text{fast}}=2\pi/k_{\text{fast}}$. This wavelength can be served as a length-scale for the fragmentation in filamentary clouds, because the fragmentation is dominated by the fastest growing eigenmode. In the absence of magnetic field, the IEOS has the smaller length-scale which is $\approx 0.78$ pc. The next smallest ones are PEOSs with $n=-4$, $n=-2$ and $n=-1$ that are followed by the GEOS with $\kappa=1$ and $\kappa=10$. The largest length-scale belongs to the MPEOS with the value $\approx 1.23$ pc. Except for $\kappa=10$, turning on the magnetic field $B=1$, would lead to drop in $\lambda_{\text{fast}}$. However, this is not the case for $B=10$ where the fragmentation length-scale of the GEOS is not only increased but also decreased more severely than the other EOSs. Also, one can see that adding the AD, leads the fragmentation length-scales to be inclined towards the pure Jeans length-scales gradually which is already observed for $\omega_{\text{fast}}$.
\section{Conclusion}\label{sec:conclusion}
It is now accepted that the filamentary MCs, play a momentous role in the first stages of star formation. According to the recent observations, the IEOS is not the best choice to describe the filamentary MCs. Softer EOSs such as the LEOS and the PEOS with the negative index are suggested by the literature to deal with this issue. In light of the new constraints imposed by the recent data, it is worthwhile to study the stability and the fragmentation of the filaments in a more accurate setting.

In this paper, we have complemented \citetalias{Rad2017}, who investigated the fragmentation of a self-gravitating filament with the IEOS which is threaded by an axial uniform magnetic field in the presence of the AD. We consider two aforementioned family of EOSs, namely the LEOS (two flavors; the GEOS and the MPEOS (see \cref{sec:EOS})) and the PEOS with negative index $(n<-1)$. We integrate the hydrostatic equation numerically. This yields us the density, the pressure and the gravitational potential profiles. Afterward, we globally perturb the fluid equations to the first order and solve the resultant ODEs using the relaxation technique. We continue with computing the dispersion relation for different EOSs, in the two magnetic field strength $B=1$ and $B=10$ and various AD regime from $\etaA = 1$ to $\etaA = 10^3$. The growth rate of the fastest growing mode $(\omega_{\text{fast}})$ can be exploited for comparison of the stability of the filament. In addition, the expected separation between clumps in a filamentary MC, can be estimated by the length-scale of the fragmentation which is predominantly determined by the wavelength of the fastest growing mode $\lambda_{\text{fast}}$. We can summarize our results as:

\begin{enumerate}
 \item In the pure Jeans instability (or equivalently when the AD is very strong), the MPEOS has the largest growth rate $\omega_{\text{fast}} \simeq 0.42$ (the shortest fragmentation time $\tau_{\text{min}} \simeq 0.41$ Myr) which followed closely by the GEOS with $\kappa=10$. This corresponds to about 25 per cent increase (decrease) in the growth rate (fragmentation time) with respect to the IEOS that has the smallest growth rate $\omega_{\text{fast}} \simeq 0.34$ (the largest fragmentation time $\tau_{\text{min}} \simeq 0.51$ Myr).
 \item The moderate magnetic field can generally increase the stability, but interestingly the degree of stabilization is very different for the two LEOSs: the MPEOS is very sensitive while the GEOS with $k=10$ is the least sensitive one.
 \item Going from the moderate magnetic field $(B=1)$ which is equivalent with $B\simeq 14.3{\upmu}$G, to the strong one $(B=10)$, the magnetic field is no longer able to effectively decrease $\omega_{\text{fast}}$ for the MPEOS and PEOSs.
 \item In the strong magnetic field without AD, the difference in $\omega_{\text{fast}}$ between all the EOSs, is negligible.
 \item Both in the moderate and the strong magnetic field, PEOSs have very similar $\omega_{\text{fast}}$.
 \item In the pure Jeans instability, the largest and the smallest fragmentation length-scales are $\simeq 1.23$ pc and $0.78$ pc which belong to the MPEOS and the IEOS respectively.
 \item In the moderate magnetic field without the AD, the fragmentation length-scale is decreased for all the EOSs specially for the MPEOS except for the GEOS with $\kappa = 10$. 
 \item In the strong magnetized medium without AD, $\lambda_{\text{fast}}$ for all the EOSs even the GEOS with $\kappa=10$ is decreased.
 \item The effect of magnetic field on the studied PEOSs is less than LEOSs. More specifically, it has the least effect on the IEOS.
\end{enumerate}

It should be noted that in our perturbation analysis, $\etaA$ is assumed to be constant, while from \cref{eq:eta_A_with_alpha} we know that it depends on the density profile. Moreover, the magnetic field can decrease the slope of the density profile at large radii \citep{FP2000I}, if it contributes in the equilibrium state. We also know that in the filamentary star forming regions, the low-density sub-filaments tend to be parallel to the magnetic field pervading the region, while the denser main filament tends to be perpendicular to the magnetic field \citep[e.g.][]{2016A&A...586A.138P}.  Another point is that in this work, filaments are not confined by the external pressure. \citet{Nagasawa87} showed that the external pressure can increase the stability of an isothermal filament by considering an infinitely hot tenuous external medium. He also showed that in this case, a uniform axial magnetic field can enhance the stability of the system by decreasing the growth rate of the instability, however, contrary to our results the magnetic field increases the critical wavelength. This is also the case for perturbations triggered in a filament which is initially in a magnetohydrostatic equilibrium state threaded by a more general helical magnetic field \citep{FP2000II}. Furthermore, \citet{Fischera2012} showed that for a filament in pressure equilibrium with the surrounding medium, a larger external pressure can lead to the considerably flatter density profiles.
In this work, we assumed that the ambipolar diffusion coefficient $\etaA$ is constant, so as we observed different density and pressure profiles could directly influence the stability properties of the filament. Considering a density dependent $\etaA$ would add to the complications.
Dealing with these problems could be the matter of next studies.

\section*{Acknowledgements}
M. Hosseinirad thanks Najme Mohammad-Salehi for useful discussions. Also, the authors would like to thank the anonymous referee for his/her helpful comments which improved the paper. This research made use of \texttt{Scipy}~\citep{scipy}, \texttt{Jupyter}~\citep{jupyter}, \texttt{Numpy}~\citep{numpy} and \texttt{Astropy}, a community-developed core Python package for Astronomy \citep{astropy}. All figures were generated using \texttt{Matplotlib}~\citep{matplotlib}. Also we have made extensive use of the NASA Astrophysical Data System Abstract Service.

\bibliography{my_paper} 

\begin{thebibliography}{}
\makeatletter
\relax
\def\mn@urlcharsother{\let\do\@makeother \do\$\do\&\do\#\do\^\do\_\do\%\do\~}
\def\mn@doi{\begingroup\mn@urlcharsother \@ifnextchar [ {\mn@doi@}
  {\mn@doi@[]}}
\def\mn@doi@[#1]#2{\def\@tempa{#1}\ifx\@tempa\@empty \href
  {http://dx.doi.org/#2} {doi:#2}\else \href {http://dx.doi.org/#2} {#1}\fi
  \endgroup}
\def\mn@eprint#1#2{\mn@eprint@#1:#2::\@nil}
\def\mn@eprint@arXiv#1{\href {http://arxiv.org/abs/#1} {{\tt arXiv:#1}}}
\def\mn@eprint@dblp#1{\href {http://dblp.uni-trier.de/rec/bibtex/#1.xml}
  {dblp:#1}}
\def\mn@eprint@#1:#2:#3:#4\@nil{\def\@tempa {#1}\def\@tempb {#2}\def\@tempc
  {#3}\ifx \@tempc \@empty \let \@tempc \@tempb \let \@tempb \@tempa \fi \ifx
  \@tempb \@empty \def\@tempb {arXiv}\fi \@ifundefined
  {mn@eprint@\@tempb}{\@tempb:\@tempc}{\expandafter \expandafter \csname
  mn@eprint@\@tempb\endcsname \expandafter{\@tempc}}}

\bibitem[\protect\citeauthoryear{{Andr{\'e}}}{{Andr{\'e}}}{2017}]{2017arXiv171001030A}
{Andr{\'e}} P.,  2017, preprint, \href
  {http://adsabs.harvard.edu/abs/2017arXiv171001030A} {} (\mn@eprint {arXiv}
  {1710.01030})

\bibitem[\protect\citeauthoryear{{Andr{\'e}}, {Men'shchikov}, {Bontemps}
  et~al.}{{Andr{\'e}} et~al.}{2010}]{Andre-2010}
{Andr{\'e}} P.,  {Men'shchikov} A.,  {Bontemps} S.,   et~al., {2010}, \mn@doi
  [{A\&A}] {10.1051/0004-6361/201014666}, {518}, {L102}

\bibitem[\protect\citeauthoryear{{Arzoumanian} et~al.,}{{Arzoumanian}
  et~al.}{2011}]{Arzoumanian2011}
{Arzoumanian} D.,  et~al., 2011, \mn@doi [\aap] {10.1051/0004-6361/201116596},
  \href {http://adsabs.harvard.edu/abs/2011A%26A...529L...6A} {529, L6}

\bibitem[\protect\citeauthoryear{{Arzoumanian}, {Andr{\'e}}, {Peretto}  \&
  {K{\"o}nyves}}{{Arzoumanian} et~al.}{2013}]{Arzoumanian2013}
{Arzoumanian} D.,  {Andr{\'e}} P.,  {Peretto} N.,   {K{\"o}nyves} V.,  2013,
  \mn@doi [\aap] {10.1051/0004-6361/201220822}, \href
  {http://adsabs.harvard.edu/abs/2013A%26A...553A.119A} {553, A119}

\bibitem[\protect\citeauthoryear{{Balbus} \& {Terquem}}{{Balbus} \&
  {Terquem}}{2001}]{BalbusTerquem2001}
{Balbus} S.~A.,  {Terquem} C.,  2001, \mn@doi [\apj] {10.1086/320452}, \href
  {http://adsabs.harvard.edu/abs/2001ApJ...552..235B} {552, 235}

\bibitem[\protect\citeauthoryear{{Bontemps} et~al.,}{{Bontemps}
  et~al.}{2010}]{Bontemps2010}
{Bontemps} S.,  et~al., 2010, \mn@doi [\aap] {10.1051/0004-6361/201014661},
  \href {http://cdsads.u-strasbg.fr/abs/2010A%26A...518L..85B} {518, L85}

\bibitem[\protect\citeauthoryear{{Burkert} \& {Hartmann}}{{Burkert} \&
  {Hartmann}}{2004}]{Burkert2004}
{Burkert} A.,  {Hartmann} L.,  2004, \mn@doi [\apj] {10.1086/424895}, \href
  {http://cdsads.u-strasbg.fr/abs/2004ApJ...616..288B} {616, 288}

\bibitem[\protect\citeauthoryear{{Camacho}, {V{\'a}zquez-Semadeni},
  {Ballesteros-Paredes}, {G{\'o}mez}, {Fall}  \& {Mata-Ch{\'a}vez}}{{Camacho}
  et~al.}{2016}]{Camacho2016}
{Camacho} V.,  {V{\'a}zquez-Semadeni} E.,  {Ballesteros-Paredes} J.,
  {G{\'o}mez} G.~C.,  {Fall} S.~M.,   {Mata-Ch{\'a}vez} M.~D.,  2016, \mn@doi
  [\apj] {10.3847/1538-4357/833/1/113}, \href
  {http://cdsads.u-strasbg.fr/abs/2016ApJ...833..113C} {833, 113}

\bibitem[\protect\citeauthoryear{{Caselli} \& {Myers}}{{Caselli} \&
  {Myers}}{1995}]{Caselli1995}
{Caselli} P.,  {Myers} P.~C.,  1995, \mn@doi [\apj] {10.1086/175825}, \href
  {http://adsabs.harvard.edu/abs/1995ApJ...446..665C} {446, 665}

\bibitem[\protect\citeauthoryear{{Chandrasekhar} \& {Fermi}}{{Chandrasekhar} \&
  {Fermi}}{1953}]{Chandra53}
{Chandrasekhar} S.,  {Fermi} E.,  1953, \mn@doi [\apj] {10.1086/145732}, \href
  {http://adsabs.harvard.edu/abs/1953ApJ...118..116C} {118, 116}

\bibitem[\protect\citeauthoryear{{Chen} \& {Ostriker}}{{Chen} \&
  {Ostriker}}{2014}]{Chen2014}
{Chen} C.-Y.,  {Ostriker} E.~C.,  2014, \mn@doi [\apj]
  {10.1088/0004-637X/785/1/69}, \href
  {http://cdsads.u-strasbg.fr/abs/2014ApJ...785...69C} {785, 69}

\bibitem[\protect\citeauthoryear{{Choi}, {Kim}  \& {Wiita}}{{Choi}
  et~al.}{2009}]{Choi2009}
{Choi} E.,  {Kim} J.,   {Wiita} P.~J.,  2009, \mn@doi [\apjs]
  {10.1088/0067-0049/181/2/413}, \href
  {http://adsabs.harvard.edu/abs/2009ApJS..181..413C} {181, 413}

\bibitem[\protect\citeauthoryear{{Cowling}}{{Cowling}}{1956}]{Cowling56}
{Cowling} T.~G.,  1956, \mn@doi [\mnras] {10.1093/mnras/116.1.114}, \href
  {http://cdsads.u-strasbg.fr/abs/1956MNRAS.116..114C} {116, 114}

\bibitem[\protect\citeauthoryear{{Dib}, {Kim}, {V{\'a}zquez-Semadeni},
  {Burkert}  \& {Shadmehri}}{{Dib} et~al.}{2007}]{Dib2007}
{Dib} S.,  {Kim} J.,  {V{\'a}zquez-Semadeni} E.,  {Burkert} A.,   {Shadmehri}
  M.,  2007, \mn@doi [\apj] {10.1086/513708}, \href
  {http://adsabs.harvard.edu/abs/2007ApJ...661..262D} {661, 262}

\bibitem[\protect\citeauthoryear{{Draine}, {Roberge}  \& {Dalgarno}}{{Draine}
  et~al.}{1983}]{Draine83}
{Draine} B.~T.,  {Roberge} W.~G.,   {Dalgarno} A.,  1983, \mn@doi [ApJ]
  {10.1086/160617}, 264, 485

\bibitem[\protect\citeauthoryear{{Elmegreen}}{{Elmegreen}}{1979}]{Elmegreen79}
{Elmegreen} B.~G.,  1979, \mn@doi [ApJ] {10.1086/157333}, 232, 729

\bibitem[\protect\citeauthoryear{{Federrath}}{{Federrath}}{2016}]{Federrath2016}
{Federrath} C.,  2016, \mn@doi [\mnras] {10.1093/mnras/stv2880}, \href
  {http://cdsads.u-strasbg.fr/abs/2016MNRAS.457..375F} {457, 375}

\bibitem[\protect\citeauthoryear{{Fiege} \& {Pudritz}}{{Fiege} \&
  {Pudritz}}{2000a}]{FP2000I}
{Fiege} J.~D.,  {Pudritz} R.~E.,  2000a, \mn@doi [\mnras]
  {10.1046/j.1365-8711.2000.03066.x}, \href
  {http://adsabs.harvard.edu/abs/2000MNRAS.311...85F} {311, 85}

\bibitem[\protect\citeauthoryear{{Fiege} \& {Pudritz}}{{Fiege} \&
  {Pudritz}}{2000b}]{FP2000II}
{Fiege} J.~D.,  {Pudritz} R.~E.,  2000b, \mn@doi [\mnras]
  {10.1046/j.1365-8711.2000.03067.x}, \href
  {http://adsabs.harvard.edu/abs/2000MNRAS.311..105F} {311, 105}

\bibitem[\protect\citeauthoryear{{Fischera} \& {Martin}}{{Fischera} \&
  {Martin}}{2012}]{Fischera2012}
{Fischera} J.,  {Martin} P.~G.,  2012, \mn@doi [\aap]
  {10.1051/0004-6361/201218961}, \href
  {http://adsabs.harvard.edu/abs/2012A%26A...542A..77F} {542, A77}

\bibitem[\protect\citeauthoryear{{Freundlich}, {Jog}  \& {Combes}}{{Freundlich}
  et~al.}{2014}]{Freundlich2014}
{Freundlich} J.,  {Jog} C.~J.,   {Combes} F.,  2014, \mn@doi [\aap]
  {10.1051/0004-6361/201323325}, \href
  {http://adsabs.harvard.edu/abs/2014A%26A...564A...7F} {564, A7}

\bibitem[\protect\citeauthoryear{{Fuller} \& {Myers}}{{Fuller} \&
  {Myers}}{1992}]{Fuller1992}
{Fuller} G.~A.,  {Myers} P.~C.,  1992, \mn@doi [\apj] {10.1086/170894}, \href
  {http://adsabs.harvard.edu/abs/1992ApJ...384..523F} {384, 523}

\bibitem[\protect\citeauthoryear{Garaud}{Garaud}{2001}]{Garaud-thesis}
Garaud P.,  2001, PhD thesis, \url
  {https://users.soe.ucsc.edu/~pgaraud/Work/thesis.pdf}

\bibitem[\protect\citeauthoryear{Gehman, Adams, Fatuzzo  \& Watkins}{Gehman
  et~al.}{1996a}]{Gehman1}
Gehman C.~S.,  Adams F.~C.,  Fatuzzo M.,   Watkins R.,  1996a, \mn@doi [ApJ]
  {10.1086/176766}, 457, 718

\bibitem[\protect\citeauthoryear{Gehman, Adams  \& Watkins}{Gehman
  et~al.}{1996b}]{Gehman2}
Gehman C.~S.,  Adams F.~C.,   Watkins R.,  1996b, \mn@doi [ApJ]
  {10.1086/178098}, 472, 673

\bibitem[\protect\citeauthoryear{{G{\'o}mez} \&
  {V{\'a}zquez-Semadeni}}{{G{\'o}mez} \&
  {V{\'a}zquez-Semadeni}}{2014}]{Gomez2014}
{G{\'o}mez} G.~C.,  {V{\'a}zquez-Semadeni} E.,  2014, \mn@doi [\apj]
  {10.1088/0004-637X/791/2/124}, \href
  {http://cdsads.u-strasbg.fr/abs/2014ApJ...791..124G} {791, 124}

\bibitem[\protect\citeauthoryear{{Gressel}, {Turner}, {Nelson}  \&
  {McNally}}{{Gressel} et~al.}{2015}]{Gressel2015}
{Gressel} O.,  {Turner} N.~J.,  {Nelson} R.~P.,   {McNally} C.~P.,  2015,
  \mn@doi [\apj] {10.1088/0004-637X/801/2/84}, \href
  {http://adsabs.harvard.edu/abs/2015ApJ...801...84G} {801, 84}

\bibitem[\protect\citeauthoryear{{Hanawa} \& {Tomisaka}}{{Hanawa} \&
  {Tomisaka}}{2015}]{Hanawa2015}
{Hanawa} T.,  {Tomisaka} K.,  2015, \mn@doi [\apj]
  {10.1088/0004-637X/801/1/11}, \href
  {http://adsabs.harvard.edu/abs/2015ApJ...801...11H} {801, 11}

\bibitem[\protect\citeauthoryear{{Hanawa}, {Kudoh}  \& {Tomisaka}}{{Hanawa}
  et~al.}{2017}]{Hanawa2017}
{Hanawa} T.,  {Kudoh} T.,   {Tomisaka} K.,  2017, \mn@doi [\apj]
  {10.3847/1538-4357/aa8b6d}, \href
  {http://adsabs.harvard.edu/abs/2017ApJ...848....2H} {848, 2}

\bibitem[\protect\citeauthoryear{{Hartmann} \& {Burkert}}{{Hartmann} \&
  {Burkert}}{2007}]{Hartmann2007}
{Hartmann} L.,  {Burkert} A.,  2007, \mn@doi [\apj] {10.1086/509321}, \href
  {http://cdsads.u-strasbg.fr/abs/2007ApJ...654..988H} {654, 988}

\bibitem[\protect\citeauthoryear{{Hennemann} et~al.,}{{Hennemann}
  et~al.}{2012}]{Hennemann2012}
{Hennemann} M.,  et~al., 2012, \mn@doi [\aap] {10.1051/0004-6361/201219429},
  \href {http://adsabs.harvard.edu/abs/2012A%26A...543L...3H} {543, L3}

\bibitem[\protect\citeauthoryear{{Heyer}, {Krawczyk}, {Duval}  \&
  {Jackson}}{{Heyer} et~al.}{2009}]{Heyer2009}
{Heyer} M.,  {Krawczyk} C.,  {Duval} J.,   {Jackson} J.~M.,  2009, \mn@doi
  [\apj] {10.1088/0004-637X/699/2/1092}, \href
  {http://adsabs.harvard.edu/abs/2009ApJ...699.1092H} {699, 1092}

\bibitem[\protect\citeauthoryear{{Hosseinirad}, {Naficy}, {Abbassi}  \&
  {Roshan}}{{Hosseinirad} et~al.}{2017}]{Rad2017}
{Hosseinirad} M.,  {Naficy} K.,  {Abbassi} S.,   {Roshan} M.,  2017, \mn@doi
  [\mnras] {10.1093/mnras/stw2820}, \href
  {http://adsabs.harvard.edu/abs/2017MNRAS.465.1645H} {465, 1645}

\bibitem[\protect\citeauthoryear{Hunter}{Hunter}{2007}]{matplotlib}
Hunter J.~D.,  2007, \mn@doi [Computing In Science \& Engineering]
  {10.1109/MCSE.2007.55}, 9, 90

\bibitem[\protect\citeauthoryear{{Inutsuka} \& {Miyama}}{{Inutsuka} \&
  {Miyama}}{1992}]{Inutsuka1992}
{Inutsuka} S.-I.,  {Miyama} S.~M.,  1992, \mn@doi [\apj] {10.1086/171162},
  \href {http://adsabs.harvard.edu/abs/1992ApJ...388..392I} {388, 392}

\bibitem[\protect\citeauthoryear{{Inutsuka}, {Inoue}, {Iwasaki}  \&
  {Hosokawa}}{{Inutsuka} et~al.}{2015}]{Inutsuka2015}
{Inutsuka} S.-i.,  {Inoue} T.,  {Iwasaki} K.,   {Hosokawa} T.,  2015, \mn@doi
  [\aap] {10.1051/0004-6361/201425584}, \href
  {http://cdsads.u-strasbg.fr/abs/2015A%26A...580A..49I} {580, A49}

\bibitem[\protect\citeauthoryear{Jones, Oliphant, Peterson  et~al.}{Jones
  et~al.}{2001}]{scipy}
Jones E.,  Oliphant T.,  Peterson P.,   et~al., 2001, {SciPy}: Open source
  scientific tools for {Python}, \url {http://www.scipy.org/}

\bibitem[\protect\citeauthoryear{{Juvela} et~al.,}{{Juvela}
  et~al.}{2012}]{Juvela2012}
{Juvela} M.,  et~al., 2012, \mn@doi [\aap] {10.1051/0004-6361/201118640}, \href
  {http://adsabs.harvard.edu/abs/2012A%26A...541A..12J} {541, A12}

\bibitem[\protect\citeauthoryear{{Klassen}, {Pudritz}  \& {Kirk}}{{Klassen}
  et~al.}{2017}]{Klassen2017}
{Klassen} M.,  {Pudritz} R.~E.,   {Kirk} H.,  2017, \mn@doi [\mnras]
  {10.1093/mnras/stw2889}, \href
  {http://cdsads.u-strasbg.fr/abs/2017MNRAS.465.2254K} {465, 2254}

\bibitem[\protect\citeauthoryear{{Klessen}, {Burkert}  \& {Bate}}{{Klessen}
  et~al.}{1998}]{Klessen1998}
{Klessen} R.~S.,  {Burkert} A.,   {Bate} M.~R.,  1998, \mn@doi [\apjl]
  {10.1086/311471}, \href {http://adsabs.harvard.edu/abs/1998ApJ...501L.205K}
  {501, L205}

\bibitem[\protect\citeauthoryear{Kluyver et~al.,}{Kluyver
  et~al.}{2016}]{jupyter}
Kluyver T.,  et~al., 2016, in ELPUB. pp 87--90

\bibitem[\protect\citeauthoryear{{K{\"o}nyves} et~al.,}{{K{\"o}nyves}
  et~al.}{2010}]{Konyves2010}
{K{\"o}nyves} V.,  et~al., 2010, \mn@doi [\aap] {10.1051/0004-6361/201014689},
  \href {http://cdsads.u-strasbg.fr/abs/2010A%26A...518L.106K} {518, L106}

\bibitem[\protect\citeauthoryear{{Larson}}{{Larson}}{1981}]{Larson1981}
{Larson} R.~B.,  1981, \mn@doi [\mnras] {10.1093/mnras/194.4.809}, \href
  {http://adsabs.harvard.edu/abs/1981MNRAS.194..809L} {194, 809}

\bibitem[\protect\citeauthoryear{{Larson}}{{Larson}}{1985}]{Larson1985}
{Larson} R.~B.,  1985, \mn@doi [\mnras] {10.1093/mnras/214.3.379}, \href
  {http://adsabs.harvard.edu/abs/1985MNRAS.214..379L} {214, 379}

\bibitem[\protect\citeauthoryear{{Lizano} \& {Shu}}{{Lizano} \&
  {Shu}}{1989}]{LizanoShu1989}
{Lizano} S.,  {Shu} F.~H.,  1989, \mn@doi [\apj] {10.1086/167640}, \href
  {http://adsabs.harvard.edu/abs/1989ApJ...342..834L} {342, 834}

\bibitem[\protect\citeauthoryear{{Mac Low}, {Norman}, {Konigl}  \&
  {Wardle}}{{Mac Low} et~al.}{1995}]{MacLow95}
{Mac Low} M.-M.,  {Norman} M.~L.,  {Konigl} A.,   {Wardle} M.,  1995, \mn@doi
  [\apj] {10.1086/175477}, \href
  {http://adsabs.harvard.edu/abs/1995ApJ...442..726M} {442, 726}

\bibitem[\protect\citeauthoryear{{Maloney}}{{Maloney}}{1988}]{Maloney1988}
{Maloney} P.,  1988, \mn@doi [\apj] {10.1086/166876}, \href
  {http://adsabs.harvard.edu/abs/1988ApJ...334..761M} {334, 761}

\bibitem[\protect\citeauthoryear{{Masson}, {Chabrier}, {Hennebelle}, {Vaytet}
  \& {Commer{\c c}on}}{{Masson} et~al.}{2016}]{Masson2016}
{Masson} J.,  {Chabrier} G.,  {Hennebelle} P.,  {Vaytet} N.,   {Commer{\c c}on}
  B.,  2016, \mn@doi [\aap] {10.1051/0004-6361/201526371}, \href
  {http://adsabs.harvard.edu/abs/2016A%26A...587A..32M} {587, A32}

\bibitem[\protect\citeauthoryear{{Matsumoto}, {Nakamura}  \&
  {Hanawa}}{{Matsumoto} et~al.}{1994}]{Matsumoto94}
{Matsumoto} T.,  {Nakamura} F.,   {Hanawa} T.,  1994, \pasj, \href
  {http://adsabs.harvard.edu/abs/1994PASJ...46..243M} {46, 243}

\bibitem[\protect\citeauthoryear{{McKee} \& {Ostriker}}{{McKee} \&
  {Ostriker}}{2007}]{McKee-2007}
{McKee} C.~F.,  {Ostriker} E.~C.,  2007, \mn@doi [\araa]
  {10.1146/annurev.astro.45.051806.110602}, \href
  {http://adsabs.harvard.edu/abs/2007ARA%26A..45..565M} {45, 565}

\bibitem[\protect\citeauthoryear{{McLaughlin} \& {Pudritz}}{{McLaughlin} \&
  {Pudritz}}{1996}]{MP96}
{McLaughlin} D.~E.,  {Pudritz} R.~E.,  1996, \mn@doi [\apj] {10.1086/177771},
  \href {http://adsabs.harvard.edu/abs/1996ApJ...469..194M} {469, 194}

\bibitem[\protect\citeauthoryear{{Men'shchikov} et~al.,}{{Men'shchikov}
  et~al.}{2010}]{Menshchikov2010}
{Men'shchikov} A.,  et~al., 2010, \mn@doi [\aap] {10.1051/0004-6361/201014668},
  \href {http://cdsads.u-strasbg.fr/abs/2010A%26A...518L.103M} {518, L103}

\bibitem[\protect\citeauthoryear{{Miesch} \& {Bally}}{{Miesch} \&
  {Bally}}{1994}]{Miesch1994}
{Miesch} M.~S.,  {Bally} J.,  1994, \mn@doi [\apj] {10.1086/174352}, \href
  {http://adsabs.harvard.edu/abs/1994ApJ...429..645M} {429, 645}

\bibitem[\protect\citeauthoryear{{Miville-Desch{\^e}nes}
  et~al.,}{{Miville-Desch{\^e}nes} et~al.}{2010}]{Miville-Deschnes2010}
{Miville-Desch{\^e}nes} M.-A.,  et~al., 2010, \mn@doi [\aap]
  {10.1051/0004-6361/201014678}, \href
  {http://cdsads.u-strasbg.fr/abs/2010A%26A...518L.104M} {518, L104}

\bibitem[\protect\citeauthoryear{{Miville-Desch{\^e}nes}, {Duc}, {Marleau},
  {Cuillandre}, {Didelon}, {Gwyn}  \& {Karabal}}{{Miville-Desch{\^e}nes}
  et~al.}{2016}]{Miville-Deschenes2016}
{Miville-Desch{\^e}nes} M.-A.,  {Duc} P.-A.,  {Marleau} F.,  {Cuillandre}
  J.-C.,  {Didelon} P.,  {Gwyn} S.,   {Karabal} E.,  2016, \mn@doi [\aap]
  {10.1051/0004-6361/201628503}, \href
  {http://adsabs.harvard.edu/abs/2016A%26A...593A...4M} {593, A4}

\bibitem[\protect\citeauthoryear{{Molinari}, {Swinyard}, {Bally}
  et~al.}{{Molinari} et~al.}{2010}]{Molinari-2010-ID667}
{Molinari} S.,  {Swinyard} B.,  {Bally} J.,   et~al., {2010}, \mn@doi [{A\&A}]
  {10.1051/0004-6361/201014659}, {518}, {L100}

\bibitem[\protect\citeauthoryear{{Nagai}, {Inutsuka}  \& {Miyama}}{{Nagai}
  et~al.}{1998}]{Nagai1998}
{Nagai} T.,  {Inutsuka} S.-i.,   {Miyama} S.~M.,  1998, \mn@doi [\apj]
  {10.1086/306249}, \href {http://adsabs.harvard.edu/abs/1998ApJ...506..306N}
  {506, 306}

\bibitem[\protect\citeauthoryear{{Nagasawa}}{{Nagasawa}}{1987}]{Nagasawa87}
{Nagasawa} M.,  1987, \mn@doi [Progress of Theoretical Physics]
  {10.1143/PTP.77.635}, \href
  {http://adsabs.harvard.edu/abs/1987PThPh..77..635N} {77, 635}

\bibitem[\protect\citeauthoryear{{Nakamura} \& {Li}}{{Nakamura} \&
  {Li}}{2008}]{Nakamura2008}
{Nakamura} F.,  {Li} Z.-Y.,  2008, \mn@doi [\apj] {10.1086/591641}, \href
  {http://cdsads.u-strasbg.fr/abs/2008ApJ...687..354N} {687, 354}

\bibitem[\protect\citeauthoryear{{Nakamura}, {Hanawa}  \& {Nakano}}{{Nakamura}
  et~al.}{1993}]{Nakamura1993}
{Nakamura} F.,  {Hanawa} T.,   {Nakano} T.,  1993, \pasj, \href
  {http://adsabs.harvard.edu/abs/1993PASJ...45..551N} {45, 551}

\bibitem[\protect\citeauthoryear{{Nakano} \& {Umebayashi}}{{Nakano} \&
  {Umebayashi}}{1986}]{NakanoUmebayashi86}
{Nakano} T.,  {Umebayashi} T.,  1986, \mn@doi [\mnras]
  {10.1093/mnras/218.4.663}, \href
  {http://adsabs.harvard.edu/abs/1986MNRAS.218..663N} {218, 663}

\bibitem[\protect\citeauthoryear{{Norman} \& {Heyvaerts}}{{Norman} \&
  {Heyvaerts}}{1985}]{Norman85}
{Norman} C.,  {Heyvaerts} J.,  1985, \aap, \href
  {http://cdsads.u-strasbg.fr/abs/1985A%26A...147..247N} {147, 247}

\bibitem[\protect\citeauthoryear{{Ntormousi}, {Hennebelle}, {Andr{\'e}}  \&
  {Masson}}{{Ntormousi} et~al.}{2016}]{Ntormousi-2016}
{Ntormousi} E.,  {Hennebelle} P.,  {Andr{\'e}} P.,   {Masson} J.,  2016,
  \mn@doi [\aap] {10.1051/0004-6361/201527400}, \href
  {http://cdsads.u-strasbg.fr/abs/2016A%26A...589A..24N} {589, A24}

\bibitem[\protect\citeauthoryear{{Oishi} \& {Mac Low}}{{Oishi} \& {Mac
  Low}}{2006}]{Oishi2006}
{Oishi} J.~S.,  {Mac Low} M.-M.,  2006, \mn@doi [\apj] {10.1086/498818}, \href
  {http://adsabs.harvard.edu/abs/2006ApJ...638..281O} {638, 281}

\bibitem[\protect\citeauthoryear{{Ostriker}}{{Ostriker}}{1964a}]{Ostriker64}
{Ostriker} J.,  1964a, \mn@doi [\apj] {10.1086/148005}, \href
  {http://adsabs.harvard.edu/abs/1964ApJ...140.1056O} {140, 1056}

\bibitem[\protect\citeauthoryear{{Ostriker}}{{Ostriker}}{1964b}]{Ostriker64b}
{Ostriker} J.,  1964b, \mn@doi [\apj] {10.1086/148057}, \href
  {http://adsabs.harvard.edu/abs/1964ApJ...140.1529O} {140, 1529}

\bibitem[\protect\citeauthoryear{{Padoan}, {Haugb{\o}lle}  \&
  {Nordlund}}{{Padoan} et~al.}{2014}]{Padoan2014}
{Padoan} P.,  {Haugb{\o}lle} T.,   {Nordlund} {\AA}.,  2014, \mn@doi [\apj]
  {10.1088/0004-637X/797/1/32}, \href
  {http://adsabs.harvard.edu/abs/2014ApJ...797...32P} {797, 32}

\bibitem[\protect\citeauthoryear{{Palmeirim} et~al.,}{{Palmeirim}
  et~al.}{2013}]{Palmeirim2013}
{Palmeirim} P.,  et~al., 2013, \mn@doi [\aap] {10.1051/0004-6361/201220500},
  \href {http://adsabs.harvard.edu/abs/2013A%26A...550A..38P} {550, A38}

\bibitem[\protect\citeauthoryear{{Panopoulou}, {Psaradaki}, {Skalidis},
  {Tassis}  \& {Andrews}}{{Panopoulou} et~al.}{2017}]{Panopoulou2017}
{Panopoulou} G.~V.,  {Psaradaki} I.,  {Skalidis} R.,  {Tassis} K.,   {Andrews}
  J.~J.,  2017, \mn@doi [\mnras] {10.1093/mnras/stw3060}, \href
  {http://adsabs.harvard.edu/abs/2017MNRAS.466.2529P} {466, 2529}

\bibitem[\protect\citeauthoryear{{Pilbratt} et~al.,}{{Pilbratt}
  et~al.}{2010}]{Pilbratt2010}
{Pilbratt} G.~L.,  et~al., 2010, \mn@doi [\aap] {10.1051/0004-6361/201014759},
  \href {http://cdsads.u-strasbg.fr/abs/2010A%26A...518L...1P} {518, L1}

\bibitem[\protect\citeauthoryear{{Planck Collaboration} et~al.,}{{Planck
  Collaboration} et~al.}{2016}]{2016A&A...586A.138P}
{Planck Collaboration} et~al., 2016, \mn@doi [\aap]
  {10.1051/0004-6361/201525896}, \href
  {http://adsabs.harvard.edu/abs/2016A%26A...586A.138P} {586, A138}

\bibitem[\protect\citeauthoryear{{Plummer}}{{Plummer}}{1911}]{Plummer1911}
{Plummer} H.~C.,  1911, \mn@doi [\mnras] {10.1093/mnras/71.5.460}, \href
  {http://cdsads.u-strasbg.fr/abs/1911MNRAS..71..460P} {71, 460}

\bibitem[\protect\citeauthoryear{{Pudritz} \& {Kevlahan}}{{Pudritz} \&
  {Kevlahan}}{2013}]{Pudritz2013}
{Pudritz} R.~E.,  {Kevlahan} N.~K.-R.,  2013, \mn@doi [Philosophical
  Transactions of the Royal Society of London Series A]
  {10.1098/rsta.2012.0248}, \href
  {http://cdsads.u-strasbg.fr/abs/2013RSPTA.37120248P} {371, 20120248}

\bibitem[\protect\citeauthoryear{{Recchi}, {Hacar}  \& {Palestini}}{{Recchi}
  et~al.}{2013}]{Recchi2013}
{Recchi} S.,  {Hacar} A.,   {Palestini} A.,  2013, \mn@doi [\aap]
  {10.1051/0004-6361/201321565}, \href
  {http://cdsads.u-strasbg.fr/abs/2013A%26A...558A..27R} {558, A27}

\bibitem[\protect\citeauthoryear{Robitaille et~al.,}{Robitaille
  et~al.}{2013}]{astropy}
Robitaille T.~P.,  et~al., 2013, Astronomy \& Astrophysics, 558, A33

\bibitem[\protect\citeauthoryear{{Sadhukhan}, {Mondal}  \&
  {Chakraborty}}{{Sadhukhan} et~al.}{2016}]{Sadhukhan2016}
{Sadhukhan} S.,  {Mondal} S.,   {Chakraborty} S.,  2016, \mn@doi [\mnras]
  {10.1093/mnras/stw837}, \href
  {http://adsabs.harvard.edu/abs/2016MNRAS.459.3059S} {459, 3059}

\bibitem[\protect\citeauthoryear{{Salmeron} \& {Wardle}}{{Salmeron} \&
  {Wardle}}{2003}]{Salmeron2003}
{Salmeron} R.,  {Wardle} M.,  2003, \mn@doi [\mnras]
  {10.1046/j.1365-8711.2003.07024.x}, \href
  {http://adsabs.harvard.edu/abs/2003MNRAS.345..992S} {345, 992}

\bibitem[\protect\citeauthoryear{{Shu}}{{Shu}}{1983}]{Shu83}
{Shu} F.~H.,  1983, \mn@doi [\apj] {10.1086/161359}, \href
  {http://adsabs.harvard.edu/abs/1983ApJ...273..202S} {273, 202}

\bibitem[\protect\citeauthoryear{{Shu}, {Adams}  \& {Lizano}}{{Shu}
  et~al.}{1987}]{Shu1987}
{Shu} F.~H.,  {Adams} F.~C.,   {Lizano} S.,  1987, \mn@doi [\araa]
  {10.1146/annurev.aa.25.090187.000323}, \href
  {http://adsabs.harvard.edu/abs/1987ARA%26A..25...23S} {25, 23}

\bibitem[\protect\citeauthoryear{{Stod{\'o}lkiewicz}}{{Stod{\'o}lkiewicz}}{1963}]{Stod63}
{Stod{\'o}lkiewicz} J.~S.,  1963, \actaa, \href
  {http://adsabs.harvard.edu/abs/1963AcA....13...30S} {13, 30}

\bibitem[\protect\citeauthoryear{{Toci} \& {Galli}}{{Toci} \&
  {Galli}}{2015}]{Toci2015}
{Toci} C.,  {Galli} D.,  2015, \mn@doi [\mnras] {10.1093/mnras/stu2168}, \href
  {http://adsabs.harvard.edu/abs/2015MNRAS.446.2110T} {446, 2110}

\bibitem[\protect\citeauthoryear{{V{\'a}zquez-Semadeni}, {G{\'o}mez},
  {Jappsen}, {Ballesteros-Paredes}, {Gonz{\'a}lez}  \&
  {Klessen}}{{V{\'a}zquez-Semadeni} et~al.}{2007}]{Vazquez-Semadeni2007}
{V{\'a}zquez-Semadeni} E.,  {G{\'o}mez} G.~C.,  {Jappsen} A.~K.,
  {Ballesteros-Paredes} J.,  {Gonz{\'a}lez} R.~F.,   {Klessen} R.~S.,  2007,
  \mn@doi [\apj] {10.1086/510771}, \href
  {http://cdsads.u-strasbg.fr/abs/2007ApJ...657..870V} {657, 870}

\bibitem[\protect\citeauthoryear{{Viala} \& {Horedt}}{{Viala} \&
  {Horedt}}{1974}]{Viala1974}
{Viala} Y.,  {Horedt} G.~P.,  1974, \aaps, \href
  {http://adsabs.harvard.edu/abs/1974A%26AS...16..173V} {16, 173}

\bibitem[\protect\citeauthoryear{Walt, Colbert  \& Varoquaux}{Walt
  et~al.}{2011}]{numpy}
Walt S. v.~d.,  Colbert S.~C.,   Varoquaux G.,  2011, Computing in Science \&
  Engineering, 13, 22

\bibitem[\protect\citeauthoryear{{Ward-Thompson} et~al.,}{{Ward-Thompson}
  et~al.}{2010}]{Ward-Thompson2010}
{Ward-Thompson} D.,  et~al., 2010, \mn@doi [\aap]
  {10.1051/0004-6361/201014618}, \href
  {http://cdsads.u-strasbg.fr/abs/2010A%26A...518L..92W} {518, L92}

\bibitem[\protect\citeauthoryear{{Wardle}}{{Wardle}}{2007}]{Wardle2007}
{Wardle} M.,  2007, \mn@doi [\apss] {10.1007/s10509-007-9575-8}, \href
  {http://adsabs.harvard.edu/abs/2007Ap%26SS.311...35W} {311, 35}

\bibitem[\protect\citeauthoryear{{Wardle} \& {Ng}}{{Wardle} \&
  {Ng}}{1999}]{WardleNg99}
{Wardle} M.,  {Ng} C.,  1999, \mn@doi [\mnras]
  {10.1046/j.1365-8711.1999.02211.x}, \href
  {http://adsabs.harvard.edu/abs/1999MNRAS.303..239W} {303, 239}

\bibitem[\protect\citeauthoryear{{Wareing}, {Pittard}, {Falle}  \& {Van
  Loo}}{{Wareing} et~al.}{2016}]{Wareing2016}
{Wareing} C.~J.,  {Pittard} J.~M.,  {Falle} S.~A.~E.~G.,   {Van Loo} S.,  2016,
  \mn@doi [\mnras] {10.1093/mnras/stw581}, \href
  {http://cdsads.u-strasbg.fr/abs/2016MNRAS.459.1803W} {459, 1803}

\bibitem[\protect\citeauthoryear{{Wurster}}{{Wurster}}{2016}]{Wurster2016}
{Wurster} J.,  2016, \mn@doi [\pasa] {10.1017/pasa.2016.34}, \href
  {http://adsabs.harvard.edu/abs/2016PASA...33...41W} {33, e041}

\bibitem[\protect\citeauthoryear{{Zhao}, {Caselli}, {Li}, {Krasnopolsky},
  {Shang}  \& {Nakamura}}{{Zhao} et~al.}{2016}]{Zhao2016}
{Zhao} B.,  {Caselli} P.,  {Li} Z.-Y.,  {Krasnopolsky} R.,  {Shang} H.,
  {Nakamura} F.,  2016, \mn@doi [\mnras] {10.1093/mnras/stw1124}, \href
  {http://adsabs.harvard.edu/abs/2016MNRAS.460.2050Z} {460, 2050}

\bibitem[\protect\citeauthoryear{{Zweibel}}{{Zweibel}}{2015}]{Zweibel2015}
{Zweibel} E.~G.,  2015, in {Lazarian} A.,  {de Gouveia Dal Pino} E.~M.,
  {Melioli} C.,  eds,  Astrophysics and Space Science Library Vol. 407,
  Magnetic Fields in Diffuse Media. p.~285,
  \mn@doi{10.1007/978-3-662-44625-6_11}

\makeatother
\end{thebibliography}
\appendix
\section{AD coefficient and equations of state in dimensionless units}\label{sec:AD_coef}
In our calculations, all quantities are transformed from cgs units to the dimensionless ones. These units are
\be [\rho]=\rho_{c},\ee
\be [t]={\sqrt{4\pi G[\rho]}}^{-1},\ee
\be [p]=p_c,\ee
\be [\mathbf{u}]=\sqrt{\dfrac{[p]}{[\rho]}},\ee
\be [r]=[t][\mathbf{u}],\ee
\be [\psi]=[\mathbf{u}]^2,\ee
\be [\mathbf{B}]=\sqrt{4\pi[p]}.\ee
It is obvious that the velocity unit is equal to the isothermal sound speed $\cs$ for the IEOS and the GEOS. For the MPEOS and the PEOS, it is assumed to be $\cs$. Using these new units, the analytical solution of the density and gravitational potential of the isothermal filament can be recast as
\begin{equation}
 \rho(r)=(1+\dfrac{r^2}{8})^{-2}
 \label{eq:rho_nond}
\end{equation}
and
\begin{equation}
 \psi(r)=2\ln(1+\dfrac{r^2}{8}).
 \label{eq:psi_nond}
\end{equation}
Furthermore, the unit of $\alpha$ can be expressed as
\begin{equation}
 [\alpha]=\dfrac{1}{4\pi (4\pi G [\rho])^{-1/2}} \frac{4\pi [\rho] [\mathbf{u}]^2}{[\mathbf{u}]^2[\rho]^{3/2}}=\sqrt{4\pi G},
 \label{eq:alpha_nond}
\end{equation}
which is $\simeq 11.465.$ Moreover, with the help of \cref{eq:eta_A_with_alpha} and \cref{eq:alpha_nond}, the unit of $\etaA$ reads
\begin{equation}
 [\etaA]=\dfrac{1}{4\pi [\alpha][\rho]^{3/2}}.
\end{equation}
This determines $\etaA$ in dimensionless units as
\begin{equation}
 \etaA \simeq 0.007 \rho_n^{-3/2}.
\end{equation}
This also transforms \cref{eq:GEOS,eq:MPEOS,eq:PEOS} to
\begin{equation}
 p = \rho + \kappa \log\,(\rho),
 \label{eq:GEOS_nond}
\end{equation}
\begin{equation}
 p = 1 + A \log\,(\rho),
 \label{eq:MPEOS_nond}
\end{equation}
\begin{equation}
 p = \rho^{\gamma_p},
 \label{eq:PEOS_nond}
\end{equation}
where $\kappa = \dfrac{p_0}{c_s^2 \rhoc}$.
\bsp	
\label{lastpage}
\end{document}